\documentclass[fleqn,10pt]{article}
\usepackage{latexsym, graphicx, epsfig, amsmath, amssymb,amsfonts}
\usepackage{natbib,amsthm,version}
\usepackage{amsbsy,bm,multirow,enumerate}
\usepackage[titletoc,page]{appendix}
\usepackage[mathscr]{eucal}
\usepackage{mathtools}
\usepackage{color}
\usepackage{subfigure}
\usepackage[utf8]{inputenc}
\usepackage[english]{babel}
\usepackage{float}
\usepackage{caption}
\usepackage{systeme}
\usepackage[affil-it]{authblk}

\newtheorem{theorem}{Theorem}[section]

\begin{document}

\title{ 
	A consistent and conservative Phase-Field model for thermo-gas-liquid-solid flows including liquid-solid phase change
	\footnote{\copyright $<$2021$>$. This manuscript version is made available under the CC-BY-NC-ND 4.0 license \url{http://creativecommons.org/licenses/by-nc-nd/4.0/}. }
	\footnote{This manuscript was accepted for publication in Journal of Computational Physics, Vol 449, Ziyang Huang, Guang Lin, Arezoo M. Ardekani, A consistent and conservative Phase-Field model for thermo-gas-liquid-solid flows including liquid-solid phase change, Page 110795, Copyright Elsevier (2021).}
} 

\author[1]{
	Ziyang Huang%
	\thanks{Email: \texttt{huan1020@purdue.edu; ziyangh@umich.edu}. Present address: Mechanical Engineering, University of Michigan, Ann Arbor, MI 48109, USA}}

\author[1,2]{
	Guang Lin%
	\thanks{Email: \texttt{guanglin@purdue.edu}; Corresponding author}}

\author[1]{
	Arezoo M. Ardekani%
	\thanks{Email: \texttt{ardekani@purdue.edu}; Corresponding author}}

\affil[1]{
	School of Mechanical Engineering, Purdue University, West Lafayette, IN 47907, USA}
\affil[2]{
	Department of Mathematics, Purdue University, West Lafayette, IN 47907, USA}

\date{}

\maketitle


\begin{abstract}
In the present study, a consistent and conservative Phase-Field model is developed to study thermo-gas-liquid-solid flows with liquid-solid phase change. The proposed model is derived with the help of the consistency conditions and exactly reduces to the consistent and conservative Phase-Field method for incompressible two-phase flows, the fictitious domain Brinkman penalization (FD/BP) method for fluid-structure interactions, and the Phase-Field model of solidification of pure material. It honors the mass conservation, defines the volume fractions of individual phases unambiguously, and therefore captures the volume change due to phase change. The momentum is conserved when the solid phase is absent, but it changes when the solid phase appears due to the no-slip condition at the solid boundary. The proposed model also conserves the energy, preserves the temperature equilibrium, and is Galilean invariant. A novel continuous surface tension force to confine its contribution at the gas-liquid interface and a drag force modified from the Carman-Kozeny equation to reduce solid velocity to zero are proposed. The issue of initiating phase change in the original Phase-Field model of solidification is addressed by physically modifying the interpolation function. The corresponding consistent scheme is developed to solve the model, and the numerical results agree well with the analytical solutions and the existing experimental and numerical data. Two challenging problems having a wide range of material properties and complex dynamics are conducted to demonstrate the capability of the proposed model.

\end{abstract}

\vspace{0.05cm}

\section{Introduction}\label{Sec Introduction}
Liquid-Solid phase change (or solidification/melting) and its interaction with the surrounding air
are ubiquitous in various natural and/or industrial processes, e.g., latent thermal energy storage (LTES) systems \citep{Shmuelietal2010,VogelThess2019}, welding \citep{Pitschenederetal1996,Chanetal1984,Zhaoetal2011,Saldi2012}, casting \citep{DantzigRappaz2016,Huangetal2018}, and additive manufacturing (AM) \citep{Panwisawasetal2017,Heetal2020,Linetal2020}. This motivates researchers to develop physical and high-fidelity models to further understand the complex physics and dynamics, and to accurately predict behaviors of materials in order to produce high-quality products. Such a problem includes many challenging factors, such as a wide range of material properties, evolution of the liquid-solid interface due to solidification/melting, deformation of the gas-liquid interface due to fluid (gas/liquid) motions and surface tension, heat transfer that drives the phase change, and fluid-structure interaction between the fluid and solid, and all these factors are coupled and can be influential. In the present study, we call the problem thermo-gas-liquid-solid flows with liquid-solid phase change. 

In spite of the complexity of the problem, it can be separated into two basic problems which are the two-phase incompressible flow and the solidification with convection. Modeling these two individual problems has been actively studied. 
For the two-phase incompressible flow, 
the front-tracking method \cite{UnverdiTryggvason1992,Tryggvasonetal2001}, 
the level-set method \cite{OsherSethian1988,Sussmanetal1994,SethianSmereka2003,Gibouetal2018}, 
the conservative level-set method \cite{OlssonKreiss2005,Olssonetal2007,ChiodiDesjardins2017}, 
the volume-of-fluid (VOF) method \cite{HirtNichols1981,ScardovelliZaleski1999,OwkesDesjardins2017}, 
the THINC method \cite{Xiaoetal2005,Iietal2012,XieXiao2017,Qianetal2018}, and 
the Phase-Field (or Diffuse-Interface) method \cite{Andersonetal1998,Jacqmin1999,Shen2011,Huangetal2020} have been developed to locate different phases.
The smoothed surface stress method \cite{Gueyffieretal1999}, 
the continuous surface force (CSF) \cite{Brackbilletal1992}, 
the ghost fluid method (GFM) \cite{Fedkiwetal1999,Lalanneetal2015}, 
the conservative and well-balanced surface tension model \cite{Abu-Al-Saud2018}, and
the Phase-Field method derived from the energy balance or the least-action principle \cite{Jacqmin1999,Yueetal2004}
have been developed to model the surface tension, and a balanced-force method \citep{Francoisetal2006} is proposed to incorporate the surface tension model to the fluid motion. Interested readers should refer to \cite{ProsperettiTryggvason2007,Tryggvasonetal2011,Shen2011,Mirjalilietal2017,Popinet2018}.
For the solidification problem with convection, one of the most popular methods is the enthalpy-porosity technique \citep{Volleretal1987,VollerPrakash1987,Brentetal1988,VollerSwaninathan1991,RoslerBruggemann2011}, which can be implemented in a fixed grid. The liquid fraction of the phase change material is algebraically related to the local temperature. Therefore, the enthalpy change due to phase change can be evaluated and becomes a source in the energy equation. A drag force proportional to the velocity is added to the momentum equation to stop the solid motion, which is the same idea as the fictitious domain Brinkman penalization (FD/BP) method for fluid-structure interactions \citep{Angotetal1999,BergmannIollo2011}. The most popularly used drag force in the enthalpy-porosity technique is modified from the Carman-Kozeny equation \citep{Carman1997} for the porous medium.
Another popular method for solidification is the Phase-Field method \citep{Boettingeretal2002,Chen2002,Echebarriaetal2004,KimKim2005,Tanetal2011,Jietal2018,Luetal2018}, where the liquid-solid interface has a small but finite thickness. Different from the enthalpy-porosity technique using an algebraic relation, the Phase-Field method introduces an additional equation to govern the evolution of the liquid fraction of the phase change material, and therefore is flexible to include more complicated physics, e.g., anisotropy and solute transport in an alloy. After coupling the Phase-Field method with the hydrodynamics, the melt convection can be modeled \citep{Nestleretal2000,Beckermannetal1999,ChenYang2019,ZhangYang2020}. 
Other methods for modeling the liquid-solid phase change are reviewed in \citep{SalcudeanAbdullah1988,Samarskiietal1993,Voller1996,HuArgyropoulos1996,Dutiletal2011,DhaidanKhodadadi2015,Sultanaetal2018}.

Most existing models for the thermo-gas-liquid-solid flows with liquid-solid phase change follow a similar procedure: The deforming gas-liquid interface is located by a certain interface capturing method and the continuous surface tension force is added, while the enthalpy-porosity technique is directly applied without any further changes to adapt to the appearance of a new phase. The volume-of-fluid method is the most popular choice and is used, e.g., in  \citep{Shmuelietal2010,Saldi2012,Kimetal2013,Yanetal2017,Panwisawasetal2017,VogelThess2019,Heetal2020}. The level-set and conservative level set methods are recently used in \citep{Yanetal2018} and \citep{Linetal2020}, respectively. Some additional physics are introduced to the models, e.g., the thermo-capillary effect \citep{Panwisawasetal2017,Yanetal2018,Heetal2020,Linetal2020} and recoil pressure \citep{Panwisawasetal2017,Heetal2020,Linetal2020}. Another recent model \citep{Zhangetal2020} follows the same strategy but instead uses the Phase-Field model in \citep{Ramirezetal2004} for anisotropic solidification and the conservative Phase-Field method \citep{ChiuLin2011} as the interface capturing method. In spite of its widespread applications, such a well-accepted modeling strategy has the following critical issues.
\textit{(i) The volume fractions of the phases are ambiguously defined.} In the models applying the enthalpy-porosity technique, the liquid fraction is meaningful only inside the phase change material, while it directly appears in the energy equation defined in the entire domain including the gas phase. As a result, the meaningless value of the liquid fraction in the gas phase is also counted in the energy equation. Another example is in \citep{Zhangetal2020}, where two liquid fractions are defined for the same liquid phase, one from the solidification model and the other from the interface capturing method. Since the solidification model and the interface capturing method have no explicit/direct connection, these two liquid fractions may inconsistently label the liquid location. 
\textit{(ii) The surface tension and drag forces can appear at wrong locations} because the volume fractions, which are not clearly defined, are needed to compute the forces. Based on the formulations, e.g., in \citep{Saldi2012,Panwisawasetal2017,Yanetal2018,Linetal2020,Heetal2020}, the surface tension at the gas-liquid interface will falsely appear at the gas-solid interface, and the drag force will falsely appear in the gas phase when the local temperature is lower than the solidus temperature. Some artificial operations need to be added but they have seldom been mentioned in the literature. An exception is in \citep{Zhangetal2020} where a bounce-back scheme near the gas-solid interface is employed since the gas-solid interface is unable to be effectively labeled by the model, but details of the bounce-back scheme are not provided. 
\textit{(iii) Physical principles can be violated, depending on material properties.} The most obvious example is the mass conservation. In, e.g., \citep{RoslerBruggemann2011,Panwisawasetal2017,Yanetal2017,Yanetal2018,Zhangetal2020,Heetal2020,Kimetal2013}, the velocity is divergence-free, implying that both the volumes of the gas and phase change material will not change. This restricts applications of the models only to problems having matched liquid and solid densities. However, the problems studied in \citep{Kimetal2013,Linetal2020} are outside that category, and therefore the volume of the phase change material needs to change in order to satisfy the mass conservation. It should be noted that the volume change in \citep{Linetal2020} is from evaporation, not solidification/melting, and the velocity is divergence-free without evaporation. As a result, the divergence-free velocity is contradicting the mass conservation. Although the studies in \citep{Shatikianetal2005,Shmuelietal2010,Hosseinizadehetal2011,VogelThess2019} captures the volume change, detail formulations, i.e., the divergence of the velocity, are not provided. Other physical principles, e.g., the momentum and energy conservation and the Galilean invariance of the models, have never been examined. 
\textit{(iv) The problem of interest often has a large density ratio}, which can be $O(10^4)$ between the gas and liquid. It has been well-known that the so-called consistent method \cite{Rudman1998,Bussmannetal2002,ChenadecPitsch2013,OwkesDesjardins2017,RaessiPitsch2012,Nangiaetal2019,Xieetal2020,Huangetal2020,Huangetal2020CAC} needs to be implemented to produce physical results for multiphase flows, while this has never been considered in the existing models for the problem of interest.

The aforementioned issues in the existing models for the problem of interest are originated in simply ``combining'', not physically ``coupling'', the models for solidification and two-phase flow. 
In the present study, those critical issues are properly addressed and a consistent and conservative Phase-Field model is developed for the thermo-gas-liquid-solid flows with liquid-solid phase change. All the dependent variables are defined in a fixed regular domain, which is convenient for numerical implementation. 

The novelty of the present study is multi-fold and the proposed model enjoys the following physical properties:
\begin{itemize}
    \item
In deriving the proposed model, several consistency conditions proposed in our previous works \citep{Huangetal2020,Huangetal2020CAC,Huangetal2020N,Huangetal2020B,Huangetal2020NPMC} are considered. Our earlier works do not include phase change or temperature variations in the fluids, and the present study is the first implementation of the consistency conditions to phase change problems, which further demonstrates their generality.
    \item 
The proposed model exactly recovers the consistent and conservative Phase-Field method for incompressible two-phase flows \citep{Huangetal2020} when the solid phase is absent, the fictitious domain Brinkman penalization (FD/BP) method for fluid-structure interactions \citep{Angotetal1999,BergmannIollo2011} when the liquid-phase is absent, and the Phase-Field model of solidification of a pure material \citep{Boettingeretal2002} when the gas phase is absent. 
    \item
The proposed model defines the volume fractions of the individual phases unambiguously and ensures their summation to be unity everywhere. The local mass conservation is strictly satisfied, from which the divergence of the velocity is non-zero, and therefore the volume change during solidification/melting is captured. 
    \item
The momentum is conserved when the solid phase is absent, and the no-slip condition at the solid boundary results in the momentum change. The energy conservation and Galilean invariance are also satisfied by the proposed model.
    \item
The momentum transport is consistent with the mass transport of the gas-liquid-solid mixture, which greatly improves the robustness of the model for large-density-ratio problems and avoids unrealistic interface deformation. 
    \item
Isothermal (or temperature equilibrium) solutions are admissible, thanks to satisfying the consistency conditions, which prevents producing any fictitious fluctuations of the temperature.
    \item
Novel continuous surface tension and drag force models are proposed, which are activated only at proper locations.
The interpolation function in the solidification model \citep{Boettingeretal2002} is modified to include the capability of initiating phase change when there is only the liquid/solid-state of the phase change material. 
\end{itemize}
These physical properties of the proposed model are independent of material properties. The proposed model is verified and its capability is demonstrated using  a consistent numerical scheme.
The proposed model includes all the basic ingredients and challenging aspects of the problem, and additional physics can be incorporated conveniently following the same framework.

The rest of the paper is organized as follows.
In Section \ref{Sec Governing equations}, the proposed model and its properties are elaborated in detail.
In Section \ref{Sec Discretizations}, the numerical procedure to solve the proposed model is introduced. 
In Section \ref{Sec Results}, various numerical tests are performed to verify the properties of the proposed model, and challenging problems are simulated to demonstrate the capability of the proposed model.
In Section \ref{Sec Conclusions}, the present study is concluded and possible future works are discussed.

\section{Governing equations}\label{Sec Governing equations}
The problem considered includes two materials, which are a gas ``$G$'' and a phase change material ``$M$'' experiencing solidification or melting. Therefore, in the entire domain $\Omega$, there are three phases: the gas phase including only ``$G$'', and the liquid and solid phases of ``$M$''. The liquid-solid phase change is driven by temperature. The part of $\Omega$ occupied by ``$G$'' is denoted by $\Omega_G$, and similarly, we use $\Omega_M$, $\Omega_M^L$, and $\Omega_M^S$ to denote the domains occupied by ``$M$'', the liquid phase, and the solid phase of ``$M$'', respectively. As a result, we have $\Omega=\Omega_G \cup \Omega_M = \Omega_G \cup \Omega_M^L \cup \Omega_M^S$. In addition, boundaries of the domains are denoted with ``$\partial$'' in front of the corresponding domains, for example, $\partial \Omega$ is the boundary of $\Omega$. Material properties of the gas phase ``$G$'' and the liquid and solid phases of ``$M$'' are assumed to be constant and denoted by $\beta_G$, $\beta_M^L$, and $\beta_M^S$, respectively, where $\beta$ can be the density $\rho$, viscosity $\mu$, specific heat $C_p$, and heat conductivity $\kappa$.

The proposed consistent and conservative model for thermo-gas-liquid-solid flows including liquid-solid phase change is elaborated in Section \ref{Sec The proposed model}. Then in Section \ref{Sec Properties}, the physical properties and the relations of the proposed model to some other multiphase models are analyzed. When the proposed model is derived in Section \ref{Sec The proposed model}, several consistency conditions will be applied so that the proposed model is able to produce physical results. The definitions of the consistency conditions are:
\begin{itemize}
    \item \textit{Consistency of reduction:} The multiphase system should be able to recover the corresponding systems including fewer phases. 
    \item \textit{Consistency of volume fraction conservation:} The phase change equation should be consistent with the
    governing equation for the volume fraction of ``$M$'', when ``$M$'' is in a fully liquid/solid-state.
	\item \textit{Consistency of mass conservation:} The mass conservation equation should be consistent with the governing equation for the volume fraction of ``$M$'', the phase change equation, and the density of the multiphase mixture. The mass flux and the divergence of the velocity in the mass conservation equation should lead to a zero mass source. 
	\item \textit{Consistency of mass and momentum transport:} The momentum flux in the momentum equation should be computed as a tensor product between the mass flux and the velocity, where the mass flux should be identical to the one in the mass conservation equation.
\end{itemize}
These consistency conditions are generalized from their correspondences for isothermal multimaterial incompressible flows \citep{Huangetal2020,Huangetal2020CAC,Huangetal2020N,Huangetal2020B,Huangetal2020NPMC} to include effects of phase change and temperature variation.

It should be noted that any thermo-gas-liquid-solid problems locally are gas-liquid, gas-solid, or liquid-solid problems and can be isothermal. Therefore, all these circumstances should be admissible by a thermo-gas-liquid-solid model in order to produce correct physical dynamics, which is emphasized by the \textit{consistency of reduction}.
To physically understand the meaning of the rest of the consistency conditions, one needs to first realize that the Phase-Field method includes relative motions of the materials, modeled as a diffusive process. The consistency conditions are general principles to incorporate the mass, momentum, and energy transports due to the relative motions of the materials into the conservation equations.

\subsection{The proposed model}\label{Sec The proposed model}

\subsubsection{The Cahn-Hilliard equation}\label{Sec Cahn-Hilliard}
The interfacial dynamics of ``$G$'' and ``$M$'' is modeled by the Cahn-Hilliard equation \citep{CahnHilliard1958} with convection:
\begin{eqnarray}\label{Eq Cahn-Hilliard}
\frac{\partial \varphi}{\partial t}
+
\nabla \cdot (\mathbf{u}\varphi)
=
\nabla \cdot (M_{\varphi} \nabla \xi_{\varphi})
+
\varphi \nabla \cdot \mathbf{u}
\quad
\mathrm{in} \quad \Omega,\\
\nonumber
\xi_{\varphi}=\lambda_{\varphi} \left( \frac{1}{\eta_\varphi^2}g'(\varphi)-\nabla^2\varphi \right),\\
\nonumber
\lambda_{\varphi}=3 \sqrt{2} \eta_{\varphi} \sigma,
\quad
g(\varphi)=\varphi^2 (1-\varphi)^2,\\
\nonumber
\mathbf{m}_{\varphi} = \mathbf{u}\varphi - M_{\varphi} \nabla \xi_{\varphi},\\
\nonumber
\mathbf{n} \cdot \nabla \xi_\varphi = \mathbf{n} \cdot \nabla \varphi = 0
\quad
\mathrm{at} \quad \partial \Omega.
\end{eqnarray}
Here, $\varphi$ is the order parameter of the Cahn-Hilliard equation and is considered as the volume fraction of ``$M$'' in $\Omega$. $\mathbf{u}$ is the velocity, whose divergence can be nonzero. $M_\varphi$ and $\xi_\varphi$ are the mobility and chemical potential of $\varphi$, respectively. $\lambda_{\varphi}$ is the mixing energy density of $\varphi$, which is proportional to both 
the thickness of ``$G$-$M$'' interface $\eta_\varphi$ and the surface tension at the gas-liquid interface $\sigma$. $g(\varphi)$ is the double-well potential, having minimums at $\varphi=0$ or $\varphi=1$, and $g'(\varphi)$ is the derivative of $g(\varphi)$ with respect to $\varphi$. $\mathbf{m}_\varphi$ is the Phase-Field flux of $\varphi$, including both the convection and diffusion fluxes in the Cahn-Hilliard equation. Unless otherwise specified, the homogeneous Neumann boundary condition is applied.

The Cahn-Hilliard equation Eq.(\ref{Eq Cahn-Hilliard}) is derived from the Ginzburg-Landau free energy functional using the $H^{-1}$ gradient flow, and the chemical potential $\xi_\varphi$ in Eq.(\ref{Eq Cahn-Hilliard}) is the functional derivative of the Ginzburg-Landau free energy functional with respect to the order parameter $\varphi$. The Cahn-Hilliard equation has been widely used in modeling two-phase incompressible flows, e.g., in \citep{Jacqmin1999,Dingetal2007,Abelsetal2012,Huangetal2020}, and its derivation has already been given in many references, e.g., in \citep{Shen2011,Fengetal2005,Yueetal2004}.

\subsubsection{The phase change equation}\label{Sec Phase change}
In the present study, we consider the Allen-Cahn Phase-Field model for the solidification of a pure material \citep{Boettingeretal2002} and the convection term is added:
\begin{eqnarray}\label{Eq Allen-Cahn}
\frac{\partial \phi}{\partial t}
+
\nabla \cdot (\mathbf{u} \phi)
=
-M_{\phi} \left(
\frac{\lambda_{\phi}}{\eta_{\phi}^2} g'(\phi)
-
\lambda_{\phi} \nabla^2 \phi
+
\frac{\rho_M^L L}{T_M} p'(\phi) (T_M-T)
\right)
+
\phi \nabla \cdot \mathbf{u}
\quad
\mathrm{in} \quad \Omega_M,\\
\nonumber
\lambda_\phi=\frac{3\sqrt{2} \rho_M^L L }{T_M} \Gamma_\phi \eta_\phi,
\quad
M_\phi =\frac{\mu_\phi \Gamma_\phi}{\lambda_\phi},\\
\nonumber
p(\phi)=\phi^3 (6\phi^2 -15\phi +10),\\
\nonumber
\mathbf{n} \cdot \nabla \phi = 0 
\quad
\mathrm{at} \quad \partial \Omega_M.
\end{eqnarray}
Notice that Eq.(\ref{Eq Allen-Cahn}) is defined in $\Omega_M$. Here, $\phi$ is the order parameter of the Allen-Cahn dynamics, and is considered as the volume fraction of the liquid phase of ``$M$'' in $\Omega_M$. $M_\phi$ is the mobility of $\phi$. $\Gamma_\phi$ and $\mu_\phi$ are the Gibbs-Thomson and linear kinetic coefficients, respectively, in the Gibbs-Thomson equation of the liquid-solid interface. $\lambda_\phi$ is the mixing energy density of $\phi$, which is related to the thickness of the liquid-solid interface $\eta_\phi$. $L$ is the latent heat of the liquid-solid phase change. $T$ is the temperature and $T_M$ is the melting temperature of the phase change. $p(\phi)$ is the interpolation function, monotonically increasing from $0$ to $1$ and having extreme points at $\phi=0$ and $\phi=1$, and $p'(\phi)$ is the derivative of $p(\phi)$ with respect to $\phi$. $g(\phi)$ is the double-well potential function defined identically to the one in Eq.(\ref{Eq Cahn-Hilliard}). Unless otherwise specified, the homogeneous Neumann boundary condition is applied. 

This solidification/melting model Eq.(\ref{Eq Allen-Cahn}) is the basis of many more complicated models including, e.g., components and/or anisotropy \citep{Boettingeretal2002,Chen2002,Jietal2018,KimKim2005,ChenYang2019,ZhangYang2020}. The model can be derived either thermodynamically from a free energy functional using the $L^2$ gradient flow or geometrically from the Gibbs-Thomson equation, and details are available in \citep{Boettingeretal2002,Beckermannetal1999,AllenCahn1979}. The first two terms in the parentheses on the right-hand side of Eq.(\ref{Eq Allen-Cahn}) models the curvature driven effect on the phase change, while they are, at the same time, competing with each other to maintain the thickness of the liquid-solid interface. The effect of the temperature is modeled by the last term in the parentheses.

Solving Eq.(\ref{Eq Allen-Cahn}) is very challenging because it is defined in $\Omega_M$ which is evolving with time. It would be more convenient if we can obtain an equivalent equation to Eq.(\ref{Eq Allen-Cahn}) but defined in $\Omega$. Therefore, the diffuse domain approach \citep{Lietal2009} is applied, and we use $\varphi$, the volume fraction of ``$M$'', as an approximation to the indicator function of $\Omega_M$ whose value is $1$ in $\Omega_M$ but $0$ elsewhere. The equivalence of Eq.(\ref{Eq Allen-Cahn}) in $\Omega$ is 
\begin{eqnarray}\label{Eq Diffuse domain approach}
\frac{\partial (\varphi \phi)}{\partial t}
+
\nabla \cdot (\mathbf{u}\varphi \phi)
=
-M_{\phi} \left(
\frac{\lambda_{\phi}}{\eta_{\phi}^2} \varphi g'(\phi)
-
\lambda_{\phi} \nabla \cdot (\varphi \nabla \phi)
+
\frac{\rho_M^L L}{T_M} \varphi p'(\phi) (T_M-T)
\right)
+
\varphi \phi \nabla \cdot \mathbf{u}
\quad
\mathrm{in} \quad \Omega,\\
\nonumber
\mathbf{n} \cdot \nabla \phi = 0
\quad
\mathrm{at} \quad \partial \Omega.
\end{eqnarray}
We directly apply the formulations of the diffuse domain approach in \citep{Lietal2009} to obtain Eq.(\ref{Eq Diffuse domain approach}) from Eq.(\ref{Eq Allen-Cahn}), and details of the approach are available in \citep{Lietal2009}. Two modifications will be performed to Eq.(\ref{Eq Diffuse domain approach}) to address the following two issues.

The first issue is about initiating the phase change. It should be noted that the terms in the parentheses on the right-hand side of Eq.(\ref{Eq Allen-Cahn}), which are the driving forces for the phase change, are nonzero only at $0 < \phi < 1$. In other words, given ``$M$'' in fully solid (liquid) state at the beginning, melting (solidification) will never happen no matter how high (low) the temperature is. This issue is originated in the definition of $p(\phi)$ in Eq.(\ref{Eq Allen-Cahn}) whose extreme points are $\phi=0$ and $\phi=1$, i.e., $p'(0)=p'(1)=0$. These extreme points are the same as the equilibrium states of $\phi$. Defining $p(\phi)$ in this way, as mentioned in \citep{Boettingeretal2002}, is an improvement from using $p(\phi)=\phi$, e.g., in \citep{Javierreetal2006}, in the sense that the equilibrium state of $\phi$ is always $0$ or $1$, independent of the temperature. However, defining $p(\phi)=\phi$ preserves the driving force from the temperature when $\phi=0$ or $\phi=1$. The equilibrium states of $\phi$ should depend on the temperature. If the temperature is larger than the melting point, the equilibrium state should be $\phi=1$ (liquid), while it should be $\phi=0$ (solid) if $T<T_M$. To achieve this property, we propose $\tilde{p}'(\phi)$ defined in Eq.(\ref{Eq Phase change}) in the present study, which combines the advantages of $p(\phi)=\phi^3 (6\phi^2 -15\phi +10)$ in \citep{Boettingeretal2002} and $p(\phi)=\phi$ in \citep{Javierreetal2006}, but avoids their disadvantages. Only when $\phi=1$ (liquid) and $T<T_M$ (undercool) or when $\phi=0$ (solid) and $T>T_M$ (overheat), $\tilde{p}'(\phi)$ is $1$, so that the effect of the temperature on the phase change is included. In other cases, $\tilde{p}'(\phi)$ is the same as $p'(\phi)$ with $p(\phi)=\phi^3 (6\phi^2 -15\phi +10)$. As a result, the equilibrium state of $\phi$ when $T<T_M$ is $\phi=0$ (solid), and it is $\phi=1$ (liquid) when $T>T_M$. 

The second issue is about the existence of fully liquid/solid state of ``$M$''. For example, given, $\phi = 1$ and $T>T_M$, we obtain $\partial \phi/\partial t=0$ from Eq.(\ref{Eq Allen-Cahn}), which implies $\phi \equiv 1$. In other words, Eq.(\ref{Eq Allen-Cahn}) admits the existence of fully liquid state of ``$M$'', when the temperature is larger than the melting temperature and there is no solid phase at the beginning. This property should be inherited by Eq.(\ref{Eq Diffuse domain approach}), while this is not the case. Using the same condition, we are unable to obtain $\phi \equiv 1$ from Eq.(\ref{Eq Diffuse domain approach}). To address this issue, we apply the \textit{consistency of volume fraction conservation}. After comparing Eq.(\ref{Eq Diffuse domain approach}) along with $\phi = 1$ and $T>T_M$ to Eq.(\ref{Eq Cahn-Hilliard}), we discover that the convection velocity $\mathbf{u}\varphi$ in Eq.(\ref{Eq Diffuse domain approach}) needs to be replaced with $\mathbf{m}_{\varphi}$. As a result, we have $\partial \varphi/\partial t + \nabla \cdot \mathbf{m}_{\varphi}$ on the left-hand side and $\varphi \nabla \cdot \mathbf{u}$ on the right-hand side, given $\phi = 1$ and $T>T_M$, and these two sides are equal to each other from the Cahn-Hilliard equation Eq.(\ref{Eq Cahn-Hilliard}). Therefore, the existence of fully liquid state of ``$M$'' is admitted by Eq.(\ref{Eq Diffuse domain approach}) after the modification. The same is also true for the solid state of ``$M$''.

After applying the above mentioned modifications to Eq.(\ref{Eq Diffuse domain approach}) to address those two issues, we obtain the phase change equation:
\begin{eqnarray}\label{Eq Phase change}
\frac{\partial (\varphi \phi)}{\partial t}
+
\nabla \cdot (\mathbf{m}_\varphi \phi)
=
-
M_{\phi} \xi_\phi 
+
\varphi \phi \nabla \cdot \mathbf{u}
\quad
\mathrm{in} \quad \Omega,\\
\nonumber
\xi_\phi 
= 
-\lambda_\phi \nabla \cdot (\varphi \nabla \phi)
+
\frac{\lambda_\phi}{\eta_\phi^2} \varphi g'(\phi)
+
\frac{\rho_M^L L}{T_M} \varphi \tilde{p}'(\phi) (T_M-T),\\
\nonumber
\lambda_\phi=\frac{3\sqrt{2} \rho_M^L L }{T_M} \Gamma_\phi \eta_\phi,
\quad
M_\phi =\frac{\mu_\phi \Gamma_\phi}{\lambda_\phi},\\
\nonumber
p(\phi)=\phi^3 (6\phi^2 -15\phi +10),
\quad
\tilde{p}'(\phi)=\left\{
\begin{array}{ll}
1, \quad \phi=0 \quad \mathrm{and} \quad T \geqslant T_M, \\
1, \quad \phi=1 \quad \mathrm{and} \quad T \leqslant T_M,\\
p'(\phi), \quad \mathrm{else}, 
\end{array}
\right.\\
\nonumber
\mathbf{m}_{\varphi \phi} = \mathbf{m}_{\varphi} \phi,\\
\nonumber
\mathbf{n} \cdot \nabla \phi = 0,
\quad
\mathrm{at} \quad \partial \Omega.
\end{eqnarray}
Here, $\xi_\phi$ is called the chemical potential of $\phi$, $\tilde{p}'(\phi)$ is modified from $p'(\phi)$ to initiate the phase change, and $\mathbf{m}_{\varphi \phi}$ is the Phase-Field flux of $(\varphi \phi)$.

\subsubsection{The volume fractions and material properties}\label{Sec Volume fractions}
Based on the Cahn-Hilliard equation Eq.(\ref{Eq Cahn-Hilliard}) and the phase change equation Eq.(\ref{Eq Phase change}), the volume fractions of the gas, liquid, and solid phases in the proposed model are defined unambiguously. The volume fraction of the gas phase (or ``$G$'') in $\Omega$ is $\alpha_G=(1-\varphi)$, and they are $\alpha_L=(\varphi \phi)$ and $\alpha_S=(\varphi -\varphi \phi)$ for the liquid and solid phases, respectively. The volume fraction of ``$M$'' in $\Omega$ is $\alpha_M=\alpha_L+\alpha_S=\varphi$. It is clear that $\alpha_G+\alpha_M=\alpha_G+\alpha_L+\alpha_S=1$ is always true in the entire domain. 

With the volume fractions of the phases in hand, the material properties of the gas-liquid-solid mixture and their fluxes are computed as
\begin{eqnarray}\label{Eq Material property}
\beta = \beta_G + (\beta_M^S-\beta_G)\varphi + (\beta_M^L-\beta_M^S) (\varphi \phi),\\
\nonumber
\mathbf{m}_{\beta} = \beta_G \mathbf{u} + (\beta_M^S-\beta_G)\mathbf{m}_{\varphi} + (\beta_M^L-\beta_M^S) \mathbf{m}_{\varphi \phi}.
\end{eqnarray}
Here, $\beta$ represents a certain material property, e.g., the density $\rho$, and $\mathbf{m}_{\varphi}$ and $\mathbf{m}_{\varphi\phi}$ are the Phase-Field fluxes defined in Eq.(\ref{Eq Cahn-Hilliard}) and Eq.(\ref{Eq Phase change}), respectively.

\subsubsection{The mass conservation}\label{Sec Mass}
To determine the mass transport of the model, the \textit{consistency of mass conservation} is applied. First, the density of the multiphase mixture is obtained following Eq.(\ref{Eq Material property})
\begin{equation}\label{Eq Denstiy}
\rho = \rho_G + (\rho_M^S-\rho_G)\varphi + (\rho_M^L-\rho_M^S) (\varphi \phi).
\end{equation}
After combining Eq.(\ref{Eq Denstiy}) with the Cahn-Hilliard equation, Eq.(\ref{Eq Cahn-Hilliard}), and the phase change equation, Eq.(\ref{Eq Phase change}), the mass of the multiphase mixture is governed by
\begin{equation}\label{Eq Mass transport}
\frac{\partial \rho}{\partial t}
+
\nabla \cdot \mathbf{m}_{\rho}
=
\rho \nabla \cdot \mathbf{u}
-
M_{\phi} (\rho_M^L-\rho_M^S)\xi_\phi,
\end{equation}
where $\mathbf{m}_\rho$ is the consistent mass flux defined in Eq.(\ref{Eq Material property}).
Finally, to obtain a zero mass source in Eq.(\ref{Eq Mass transport}), the divergence of the velocity should satisfy
\begin{eqnarray}\label{Eq Divergence}
\nabla \cdot \mathbf{u}
=
\frac{M_\phi (\rho_M^L-\rho_M^S)}{\rho} \xi_\phi.
\end{eqnarray}
Eq.(\ref{Eq Divergence}) illustrates the volume change due to the liquid-solid phase change.
As a result, the mass conservation equation of the model is
\begin{equation}\label{Eq Mass}
\frac{\partial \rho}{\partial t}
+
\nabla \cdot \mathbf{m}_{\rho}
=
0.
\end{equation}
Therefore, the mass of the gas-liquid-solid mixture is locally conserved, even though the phase change happens. This is achieved at the expense of changing the volume of ``$M$'', as indicated in Eq.(\ref{Eq Divergence}). On the other hand, if the densities of the liquid and solid phases of ``$M$'' are the same, the volume of ``$M$'' remains the same, and therefore the divergence of the velocity is zero, which can also be seen in Eq.(\ref{Eq Divergence}).
It should be noted that the mass conservation equation Eq.(\ref{Eq Mass}) is not an independent equation in the model. Instead, it is derived from Eq.(\ref{Eq Material property}) and Eq.(\ref{Eq Divergence}), as elaborated in this section. Therefore, we don't need to explicitly solve the mass conservation equation Eq.(\ref{Eq Mass}).

\subsubsection{The momentum equation}\label{Sec Momentum}
The motion of the phases (or materials) is governed by the momentum equation:
\begin{eqnarray}\label{Eq Momentum}
\frac{\partial (\rho \mathbf{u})}{\partial t}
+
\nabla \cdot (\mathbf{m}_\rho \otimes \mathbf{u})
=
-
\nabla P
+
\nabla \cdot [\mu (\nabla \mathbf{u}+\nabla \mathbf{u}^T)]
+
\rho \mathbf{g}
+
\mathbf{f}_s
+
\mathbf{f}_d,\\
\nonumber
\mathbf{f}_s=\phi \xi_\varphi \nabla \varphi,
\quad
\mathbf{f}_d= A_d (\mathbf{u}_S-\mathbf{u}),
\quad
A_d = C_d \frac{\alpha_S^2}{(1-\alpha_S)^3+e_d}.
\end{eqnarray}
Here, $P$ is the pressure, and $\mathbf{g}$ is the gravity. As all the gas-liquid, gas-solid, and liquid-solid interfaces are immersed in $\Omega$, their effects are modeled as volumetric forces, i.e., $\mathbf{f}_s$ and $\mathbf{f}_d$, in the momentum equation Eq.(\ref{Eq Momentum}). $\mathbf{f}_s$ is the surface tension force, modeling the surface tension at the gas-liquid interface, while $\mathbf{f}_d$ is the drag force, modeling the no-slip boundary condition at the gas-solid and liquid-solid interfaces by enforcing $\mathbf{u}=\mathbf{u}_S$ in the solid phase. $A_d$ is the drag coefficient of $\mathbf{f}_d$, and $C_d$ and $e_d$ are model parameters of $A_d$. $\mathbf{u}_S$ is the given solid velocity and is set to be zero unless otherwise specified.
Note that the momentum is transported with the consistent mass flux $\mathbf{m}_\rho$ that also appears in the mass conservation equation Eq.(\ref{Eq Mass}). This follows the \textit{consistency of mass and momentum transport}, and is essential to obtain the kinetic energy conservation (when only the pressure gradient is present on the right-hand side of Eq.(\ref{Eq Momentum})) and the Galilean invariance, as analyzed in \citep{Huangetal2020}.

\begin{figure}[!t]
	\centering
	\includegraphics[scale=.4]{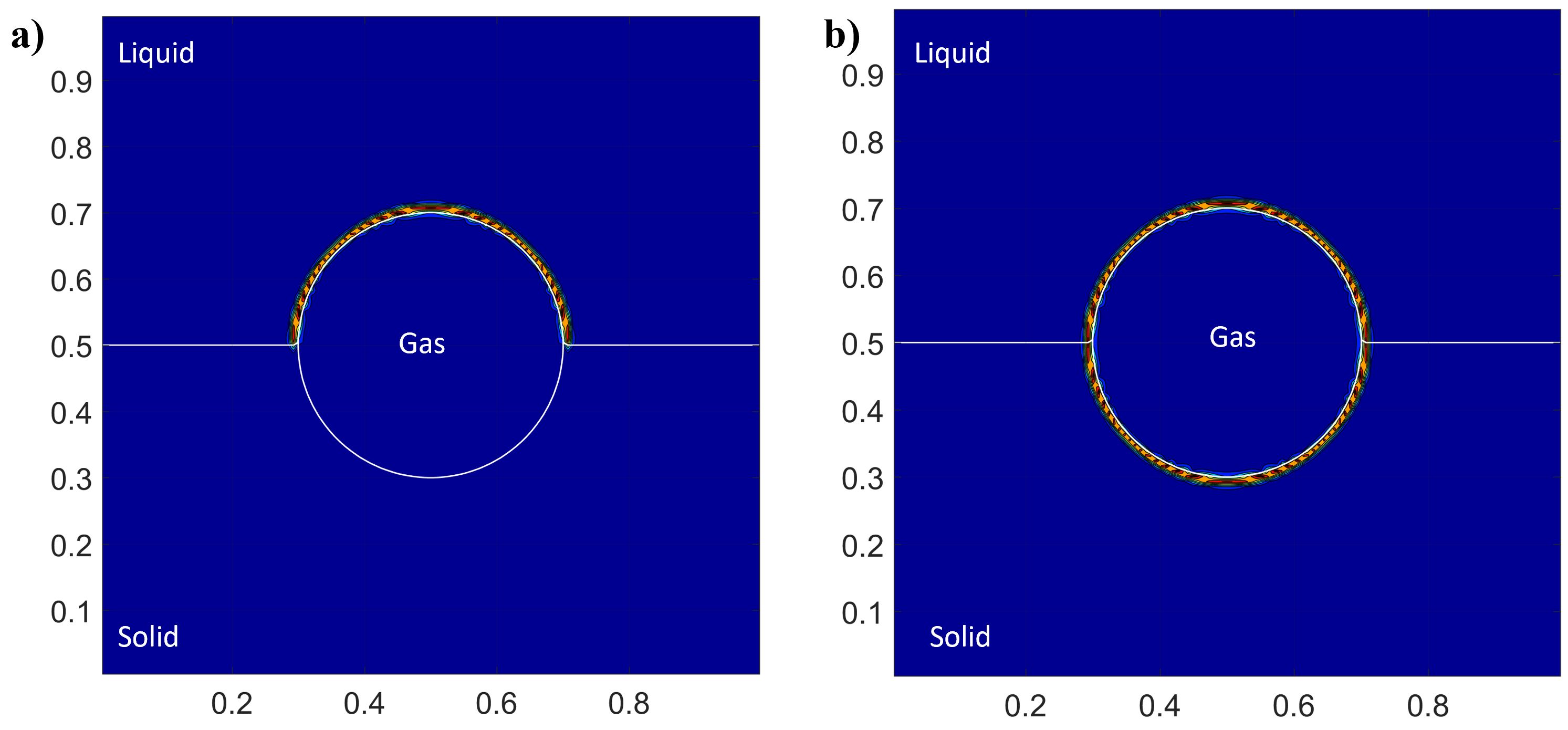}\\
	\caption{Magnitude of the surface tension force $|\mathbf{f}_s|$. a) The proposed surface tension force $\mathbf{f}_s=\phi \xi_\varphi \nabla \varphi$, b) The original surface tension force $\mathbf{f}_s=\xi_\varphi \nabla \varphi$. The proposed surface tension force shown in a) removes its contribution at the gas-solid interface, compared to the original formulation shown in b).\label{Fig Fs}}
\end{figure}

To model the surface tension at the gas-liquid interface, we apply the commonly used Phase-Field formulation $\xi_\varphi \nabla \varphi$, which can be derived from either the energy balance \cite{Jacqmin1999,Huangetal2020CAC} or the least-action principle \cite{Yueetal2004,Shen2011}. However, $\xi_\varphi \nabla \varphi$ is activated at all ``$G$-$M$'' interfaces, including both the gas-liquid and gas-solid interfaces. To remove its contribution at the gas-solid interface, $\xi_\varphi \nabla \varphi$ is multiplied by $\phi$ so that the surface tension force remains to be $\xi_\varphi \nabla \varphi$ at the gas-liquid interface and smoothly reduces to zero away from it. As a result, we obtain $\mathbf{f}_s=\phi \xi_\varphi \nabla \varphi$ in Eq.(\ref{Eq Momentum}). 
To more clearly illustrate the distribution of the surface tension force, we consider a gas bubble at the center of a unit domain with a radius $0.2$, the bottom half of which is in contact with the solid phase while the upper half is in contact with the liquid phase. Fig.\ref{Fig Fs} schematically shows the magnitude of the surface tension forces. We will further investigate the surface tension force in Section \ref{Sec Fources}. It should be noted that the thermo-capillary effect has not been considered in the present study and the surface tension $\sigma$ is treated as a constant.  

To model the no-slip boundary condition at both the gas-solid and liquid-solid interfaces, the velocity inside the solid phase needs to be the given value $\mathbf{u}_S$. We add a drag force, formulated as $\mathbf{f}_d=A_d (\mathbf{u_S}-\mathbf{u})$, to the momentum equation Eq.(\ref{Eq Momentum}), where the drag coefficient $A_d$ can be considered as a penalty coefficient that enforces $\mathbf{u}=\mathbf{u}_S$ inside the solid phase. $A_d$ should depend on the volume fraction of the solid phase in such a way that it is zero away from the solid phase and increases to a large enough value that overwhelms the inertia and viscous effects inside the solid phase. As a result, around the gas-solid and liquid-solid interfaces, the momentum equation Eq.(\ref{Eq Momentum}) reduces to Darcy’s law \citep{Volleretal1987}. Voller and Prakash \citep{VollerPrakash1987} modified the well-known Carman-Kozeny equation \citep{Carman1997}, by adding a small constant, denoted by $e_d$ here, at the denominator to avoid division by zero, and obtained $A_d=C_d \frac{(1-\phi)^2}{\phi^3+e_d}$ in $\Omega_M$, where $C_d$ should be a large number and $\phi$ is the volume fraction of the liquid phase of ``$M$'' in $\Omega_M$ as a reminder. Such a definition of $A_d$ has been popularly used, e.g., in \citep{Brentetal1988,Shmuelietal2010,RoslerBruggemann2011,Panwisawasetal2017,VogelThess2019}. It should be noted that Voller and Prakash \citep{VollerPrakash1987} did not consider the gas phase and formulated $A_d$ in terms of the volume fraction of the liquid phase. Therefore, their formulation is only applicable in $\Omega_M$. To obtain $A_d$ in $\Omega$, we directly use the volume fraction of the solid phase in $\Omega$ as the dependent variable of $A_d$, i.e., $A_d=C_d \frac{\alpha_S^2}{(1-\alpha_S)^3+e_d}$, as shown in Eq.(\ref{Eq Momentum}). 
Another popular option of enforcing $\mathbf{u}=0$ is to assign a large viscosity inside the solid phase, e.g., in \citep{Volleretal1987,Yanetal2017,Yanetal2018}. Although it looks simpler, our numerical tests in Section \ref{Sec Fources} show that this is not an effective choice and adding a drag force is recommended.

\subsubsection{The energy equation}\label{Sec Energy}
The enthalpy of the multiphase mixture is $\int_\Omega \left( (\rho C_p) T+ \rho_M^L L (\varphi \phi) \right) d\Omega$, and only the heat conductivity is considered in the present study. As a result, we obtain the following energy equation:
\begin{equation}\label{Eq Energy2}
\frac{\partial ((\rho C_p)T)}{\partial t}
+
\nabla \cdot ( \mathbf{u}(\rho C_p) T )
+
\rho_M^L L \left[
\frac{\partial (\varphi \phi)}{\partial t}
+
\nabla \cdot (\mathbf{u} \varphi \phi)
\right]
=
\nabla \cdot (\kappa \nabla T)+
Q_T.
\end{equation}
Here, $(\rho C_p)$ and $\kappa$ denote the volumetric heat and the heat conductivity, respectively, and they are computed from Eq.(\ref{Eq Material property}). It should be noted that $(\rho C_p)$ here is considered as a single quantity, which is different from the product of $\rho$ and $C_p$ that are computed individually from Eq.(\ref{Eq Material property}). $Q_T$ is the heat source, and is neglected unless otherwise specified. 

The terms in the bracket on the left-hand side of Eq.(\ref{Eq Energy2}) represents the effect of the phase change to the enthalpy. Therefore, without the phase change, i.e., $\phi \equiv 1$ and $T>T_M$ or $\phi \equiv 0$ and $T<T_M$, these terms should disappear. To achieve this goal, $\nabla \cdot (\mathbf{u} \varphi \phi)$ in Eq.(\ref{Eq Energy2}) needs to be replaced by $\nabla \cdot (\mathbf{m}_\varphi \phi)$ or $\nabla \cdot (\mathbf{m}_{\varphi \phi})$, due to the \textit{consistency of volume fraction conservation}. As a result, the terms in the bracket in Eq.(\ref{Eq Energy2}) after the modification are identical to those on the left-hand side of Eq.(\ref{Eq Phase change}).

Further, the physical energy equation should admit isothermal (or temperature equilibrium) solutions when the phase change does not happen. Plugging $T \equiv T_0$ in Eq.(\ref{Eq Energy2}), we obtain $T_0 \left[ \frac{\partial (\rho C_p)}{\partial t}+\nabla \cdot ( \mathbf{u}(\rho C_p))\right]$ on the left-hand side and $0$ on the right-hand side. However, $\left[ \frac{\partial (\rho C_p)}{\partial t}+\nabla \cdot ( \mathbf{u}(\rho C_p))\right]$ is not zero with $(\rho C_p)$ computed from Eq.(\ref{Eq Material property}). Therefore, the isothermal solution, i.e., $T \equiv T_0$, is not admissible by Eq.(\ref{Eq Energy2}). Following the derivation from the \textit{consistency of mass conservation} in Section \ref{Sec Mass} and replacing $\rho$ with $(\rho C_p)$, we can show that $\left[ \frac{\partial (\rho C_p)}{\partial t}+\nabla \cdot \mathbf{m}_{(\rho C_p)} \right]$ is zero, where $\mathbf{m}_{(\rho C_p)}$ is the flux of volumetric heat defined in Eq.(\ref{Eq Material property}) as well, by noticing that the velocity is divergence-free when there is no phase change. Therefore, on the left-hand side of Eq.(\ref{Eq Energy2}), $\nabla \cdot (\mathbf{u}(\rho C_p)T)$ needs to be replaced by $\nabla \cdot (\mathbf{m}_{(\rho C_p)}T)$. 
Combining the above modifications to Eq.(\ref{Eq Energy2}), we obtain the consistent energy equation:
\begin{equation}\label{Eq Energy}
\frac{\partial ((\rho C_p) T)}{\partial t}
+
\nabla \cdot (\mathbf{m}_{(\rho C_p)} T)
+
\rho_M^L L \left(
\frac{\partial (\varphi \phi)}{\partial t}
+
\nabla \cdot \mathbf{m}_{\varphi \phi}
\right)
=
\nabla \cdot (\kappa \nabla T)
+
Q_T.
\end{equation}

\subsection{Properties}\label{Sec Properties}
Eq.(\ref{Eq Cahn-Hilliard}), Eq.(\ref{Eq Phase change}), Eq.(\ref{Eq Material property}), Eq.(\ref{Eq Divergence}), Eq.(\ref{Eq Momentum}), and Eq.(\ref{Eq Energy}) complete the consistent and conservative model for thermo-gas-liquid-solid flows including liquid-solid phase change, and the proposed model honors many physical properties. From Eq.(\ref{Eq Mass}), Eq.(\ref{Eq Momentum}), and Eq.(\ref{Eq Energy}), it is obvious that the mass and enthalpy of the multiphase mixture are conserved, and the momentum (neglecting the gravity) is conserved without the appearance of the solid phase, by noticing that $\mathbf{f}_s=\xi_\varphi \nabla \varphi$ is equivalent to $\nabla \cdot (-\lambda_\varphi \nabla \varphi \otimes \nabla \varphi)$, see \citep{Jametetal2002,Jacqmin1999,Shen2011,Huangetal2020N}, and $\mathbf{f}_d=\mathbf{0}$, in this circumstance. When there is the solid phase, the momentum is not necessarily conserved due to the no-slip boundary condition at the solid boundary. 
The Galilean invariance is also satisfied by the proposed model. The proof is straightforward, using the Galilean transformation, and examples are available in \citep{Huangetal2020,Huangetal2020NPMC}. The subtle part of the proof is related to the left-hand side of the momentum equation, and we need to emphasize that the consistency conditions are playing a critical role there. 

More importantly, the proposed model is reduction consistent with (i) the isothermal consistent and conservative Phase-Field method for two-phase incompressible flows in \citep{Huangetal2020} when the solid phase is absent and the initial homogeneous temperature is larger than the melting temperature, (ii) the isothermal fictitious domain Brinkman penalization (FD/BP) method for fluid-structure interactions in \citep{Angotetal1999,BergmannIollo2011} when the liquid phase is absent and the initial homogeneous temperature is lower than the melting temperature, and (iii) the Phase-Field model of solidification in \citep{Boettingeretal2002} when both the gas phase and the flow are absent and the material properties of the liquid and solid phases are matched (except the thermal conductivities).

\begin{theorem}\label{Theorem Two-Phase}
The proposed model in Section \ref{Sec The proposed model} is consistent with the isothermal consistent and conservative Phase-Field method for two-phase incompressible flows in \citep{Huangetal2020}.
\end{theorem}
\begin{proof}\label{Proof Two-Phase}
Given $\phi = 1$ and $T = T_0 > T_M$ at $t=0$, as already analyzed in Section \ref{Sec Phase change} and Section \ref{Sec Energy}, we obtain $\phi = 1$ and therefore $\mathbf{m}_{\varphi \phi}=\mathbf{m}_{\varphi}$ and $\xi_\phi=0$ from Eq.(\ref{Eq Phase change}), and $T=T_0$ from Eq.(\ref{Eq Energy}), at $\forall t>0$. As a result, the velocity is divergence-free from Eq.(\ref{Eq Divergence}), and the last term on the right-hand side of the Cahn-Hilliard equation Eq.(\ref{Eq Cahn-Hilliard}) vanishes. From Eq.(\ref{Eq Material property}), we obtain $\beta = \beta_G + (\beta_M^L-\beta_G)\varphi$ and $\mathbf{m}_{\beta} = \beta_G \mathbf{u} + (\beta_M^L-\beta_G)\mathbf{m}_{\varphi}$, showing that the contribution of $\beta_M^S$ disappears. Finally in Eq.(\ref{Eq Momentum}), $\mathbf{f}_s$ becomes $\xi_\varphi \nabla \varphi$, and $\mathbf{f}_d=\mathbf{0}$ due to $\alpha_S=(\varphi-\varphi \phi)=0$. Therefore, the temperature remains its homogeneous initial value, and the simplified system from the proposed model in Section \ref{Sec The proposed model} with the given condition is equivalent to the consistent and conservative Phase-Field method for two-phase incompressible flows in \citep{Huangetal2020}, by noticing that $\tilde{\varphi}=2\varphi-1$ is the order parameter of the Cahn-Hilliard equation in \citep{Huangetal2020}.
\end{proof}

\begin{theorem}\label{Theorem FSI}
The proposed model in Section \ref{Sec The proposed model} is consistent with the isothermal fictitious domain Brinkman penalization (FD/BP) method for fluid-structure interactions in \citep{Angotetal1999,BergmannIollo2011}.
\end{theorem}
\begin{proof}\label{Proof FSI}
Given $\phi = 0$ and $T = T_0 < T_M$ at $t=0$, as already analyzed in Section \ref{Sec Phase change} and Section \ref{Sec Energy}, we obtain $\phi = 0$ and therefore $\mathbf{m}_{\varphi \phi}=\mathbf{0}$ and $\xi_\phi=0$ from Eq.(\ref{Eq Phase change}), and $T=T_0$ from Eq.(\ref{Eq Energy}), at $\forall t>0$. As a result, the velocity is divergence-free from Eq.(\ref{Eq Divergence}), and the last term on the right-hand side of the Cahn-Hilliard equation Eq.(\ref{Eq Cahn-Hilliard}) vanishes. In Eq.(\ref{Eq Momentum}), $\mathbf{f}_s$ becomes $\mathbf{0}$, and $\mathbf{f}_d=A_d(\mathbf{u}_S-\mathbf{u})$ with $\alpha_S=\varphi$. In this case, the gas phase should be understood as an arbitrary incompressible fluid and the solid phase represents the fictitious domain, where the material properties are the same as the fluid ones without loss of generality. As a result, we obtain $\beta = \beta_G$ and $\mathbf{m}_{\beta} = \beta_G \mathbf{u}$ from Eq.(\ref{Eq Material property}). Therefore, the temperature remains its initial homogeneous value, and the simplified system from the proposed model in Section \ref{Sec The proposed model} with the given condition is equivalent to the FD/BP method for fluid-structure interactions in \citep{Angotetal1999,BergmannIollo2011}. Notice that $A_d$ is defined proportional to $\alpha_S$ in \citep{Angotetal1999,BergmannIollo2011}, different from the one in Eq.(\ref{Eq Momentum}), and the level-set method, instead of the Cahn-Hilliard equation, is used in \citep{BergmannIollo2011} for the volume fraction of the fictitious domain (or the solid phase).
\end{proof}

\begin{theorem}\label{Theorem Solidification}
The proposed model in Section \ref{Sec The proposed model} is consistent with the Phase-Field model of solidification in \citep{Boettingeretal2002}.
\end{theorem}
\begin{proof}\label{Proof Solidification}
Given $\varphi = 1$ and $\mathbf{u}=\mathbf{0}$ at $t=0$, we have $\xi_\varphi = 0$ and therefore $\partial \varphi/\partial t=0$ from Eq.(\ref{Eq Cahn-Hilliard}), which implies $\varphi=1$ and $\mathbf{m}_{\varphi}=\mathbf{u}$ at $\forall t>0$. Further requiring that the material properties of the liquid and solid phases of ``$M$'' are identical, we obtain $\nabla \cdot \mathbf{u}=0$ and $\partial \mathbf{u}/{\partial t}=\mathbf{0}$ from Eq.(\ref{Eq Divergence}) and Eq.(\ref{Eq Momentum}), respectively. Therefore, we obtain $\mathbf{u}=\mathbf{0}$ at $\forall t>0$, and all the convection terms are dropped. Putting all these to the phase change equation Eq.(\ref{Eq Phase change}) and the energy equation Eq.(\ref{Eq Energy}), they become the same as those in \citep{Boettingeretal2002}, except that $p'(\phi)$ is replaced with $\tilde{p}'(\phi)$ in the present study.
\end{proof}

\textit{\textbf{Remark:}
In the proofs of Theorem \ref{Theorem Two-Phase}, Theorem \ref{Theorem FSI}, and Theorem \ref{Theorem Solidification}, the given conditions are assumed to be true at $t=0$ in the entire domain for convenience. Actually, we only need those conditions to be true locally at any moment, and Theorem \ref{Theorem Two-Phase}, Theorem \ref{Theorem FSI}, and Theorem \ref{Theorem Solidification} will again be valid. In other words, the proposed model in Section \ref{Sec The proposed model} will automatically reduce to the corresponding multiphase models whenever one of the phases is locally absent. 
}

\section{Discretization of the governing equations}\label{Sec Discretizations}
The numerical procedure to solve the proposed model in Section \ref{Sec The proposed model} is introduced in this section. The differential operators are discretized with the conservative finite difference method as those in \citep{Huangetal2020}, such that the convection terms are discretized by the 5th-order WENO scheme \cite{JiangShu1996}, while the divergence, gradient, and Laplacian operators are approximated by the 2nd-order central difference \citep{FerzigerPeric2001}. These discrete operators have been carefully verified in various studies \citep{Huangetal2020,Huangetal2020CAC,Huangetal2020N,Huangetal2020B,Huangetal2020NPMC}. Following the notations in \citep{Huangetal2020}, the discrete operators are denoted by $\tilde{(\cdot)}$, and time levels are indicated by superscript.
We use $\frac{\gamma_t f^{n+1}-\hat{f}}{\Delta t}$ to denote the discretization of the time derivative $\frac{\partial f}{\partial t}$, and $f^{*,n+1}$ to denote the approximation of $f^{n+1}$ from previous time levels. Here, $\gamma_t=1$, $\hat{f}=f^n$, and $f^{*,n+1}=f^n$ in the 1st-order case, while they are $\gamma_t=1.5$, $\hat{f}=2f^{n}-0.5f^{n-1}$, and $f^{*,n+1}=2f^{n}-f^{n-1}$ in the 2nd order case. Unless otherwise specified, we use the 2nd-order scheme.
The major concern is reproducing the physical connections among the governing equations, discussed in Section \ref{Sec The proposed model}, at the discrete level, following the consistency conditions. 

First, the Cahn-Hilliard equation Eq.(\ref{Eq Cahn-Hilliard}) is solved with the convex splitting scheme in \citep{DongShen2012,Huangetal2020}, along with the consistent and conservative boundedness mapping \citep{Huangetal2020CAC}. Then, the fully-discretized Cahn-Hilliard equation is recovered and rearranged to be
\begin{equation}\label{Eq Cahn-Hilliard discrete}
\frac{\gamma_t \varphi^{n+1}-\hat{\varphi}}{\Delta t}
+
\tilde{\nabla} \cdot \tilde{\mathbf{m}}_{\varphi}
=
\varphi^{*,n+1} \tilde{\nabla} \cdot \mathbf{u}^{*,n+1}.
\end{equation}
Therefore, we obtain $\varphi^{n+1}$ and the discrete Phase-Field flux $\tilde{\mathbf{m}}_{\varphi}$ after solving the Cahn-Hilliard equation. 

Second, we proceed to solve the phase change equation Eq.(\ref{Eq Phase change}). To preserve the \textit{consistency of volume fraction conservation} on the discrete level, it should be noted that $\varphi^{n+1}$ and $\tilde{\mathbf{m}}_{\varphi}$ in Eq.(\ref{Eq Cahn-Hilliard discrete}) are inputs to solve the phase change equation. The fully-discretized phase change equation is
\begin{eqnarray}\label{Eq Phase change discrete}
\frac{ \gamma_t (\varphi^{n+1} \phi^{n+1})-\widehat{(\varphi \phi)}}{\Delta t}
+
\tilde{\nabla} \cdot (\tilde{\mathbf{m}}_\varphi \tilde{\phi}^{*,n+1})
=
-
M_{\phi} \tilde{\xi}_\phi 
+
\varphi^{*,n+1} \phi^{*,n+1} \tilde{\nabla} \cdot \mathbf{u}^{*,n+1},\\
\nonumber
\tilde{\xi}_\phi 
= 
-\lambda_\phi \tilde{\nabla} \cdot (\overline{\varphi^{n+1}} \tilde{\nabla} \phi^{n+1})
+
\frac{\lambda_\phi}{\eta_\phi^2} \varphi^{n+1} \tilde{g}'(\phi^{n+1})
+
\frac{\rho_M^L L}{T_M} \varphi^{n+1} \tilde{p}'(\phi^{*,n+1}) (T_M-T^{*,n+1}),\\
\nonumber
\tilde{\mathbf{m}}_{\varphi \phi} = \tilde{\mathbf{m}}_{\varphi} \tilde{\phi}^{*,n+1},
\end{eqnarray}
where $\tilde{\phi}$ represents the WENO reconstruction, $\overline{\varphi}$ denotes the linear interpolation, and $\tilde{g}'(\phi^{n+1})$ is linearized $g'(\phi^{n+1})$ around $\phi^{*,n+1}$ from Taylor expansion.

After solving Eq.(\ref{Eq Cahn-Hilliard discrete}) and Eq.(\ref{Eq Phase change discrete}), $(\rho C_p)$, $\tilde{\mathbf{m}}_{(\rho C_p)}$, and $\kappa$ are obtained from Eq.(\ref{Eq Material property}), noticing that $\mathbf{u}^{*,n+1}$ is applied to Eq.(\ref{Eq Material property}) due to the \textit{consistency of reduction}, see \citep{Huangetal2020N}. Then, the temperature is updated from the following fully-discretized energy equation:
\begin{eqnarray}\label{Eq Energy discrete}
\frac{\gamma_t(\rho C_p)^{n+1} T^{n+1}-\widehat{(\rho C_p) T}}{\Delta t}
+
\tilde{\nabla} \cdot (\tilde{\mathbf{m}}_{(\rho C_p)} \tilde{T}^{*,n+1})
+
\rho_M^L L \left[
\frac{\gamma_t (\varphi^{n+1} \phi^{n+1})-\widehat{(\varphi \phi)} }{\Delta t}
+
\tilde{\nabla} \cdot \tilde{\mathbf{m}}_{\varphi \phi}
\right]\\
\nonumber
=
\tilde{\nabla} \cdot (\overline{\kappa^{n+1}} \tilde{\nabla} T^{n+1})
+
Q_T^{n+1},
\end{eqnarray}
where $\tilde{T}$ represents the WENO reconstruction and $\overline{\kappa}$ denotes the linear interpolation.
It should be noted that the terms in the bracket in Eq.(\ref{Eq Energy discrete}) are identical to the left-hand side of the fully-discretized phase change equation in Eq.(\ref{Eq Phase change discrete}). This numerical correspondence is consistent with the derivation in Section \ref{Sec Energy}.

Finally, the momentum equation Eq.(\ref{Eq Momentum}) is solved, majorly based on the 2nd-order projection scheme on a collocated grid \citep{Huangetal2020}, which has been carefully analyzed and successfully applied to two- and multi-phase problems \cite{Huangetal2020,Huangetal2020N,Huangetal2020CAC}. 
Again, $\rho$, $\tilde{\mathbf{m}}_\rho$, and $\mu$ in the momentum equation are directly computed from Eq.(\ref{Eq Material property}) and the surface tension force $\mathbf{f}_s^{n+1}$ is computed from its definition in Eq.(\ref{Eq Momentum}) with the balanced-force method \citep{Huangetal2020,Francoisetal2006}, while the drag force $\mathbf{f}_d$ is treated implicitly. As mentioned in Section \ref{Sec Momentum}, $\mathbf{f}_d$ should be predominant over either the inertial or viscous effect inside the solid phase. In other words, from Eq.(\ref{Eq Momentum}), $A_d|_{\alpha_S=1}=C_d/e_d$ should be much larger than $\rho/\Delta t$ (inertia) or $\mu/h^2$ (viscous). Here $h$ denotes the grid size. To achieve this goal, we set $C_d$ as $(\rho^{n+1}/\Delta t+\mu^{n+1}/h^2)$, and $e_d$ is fixed to be $10^{-3}$. Therefore, $\mathbf{f}_d$ will always be thousand times larger than both the inertial and viscous effects inside the solid phase, regardless of the numerical setup or material properties.
The divergence of the velocity at the new time level, which appears in the projection scheme, is determined from Eq.(\ref{Eq Divergence}), i.e., 
\begin{eqnarray}\label{Eq Divergence discrete}
\tilde{\nabla} \cdot \mathbf{u}^{n+1}
=
\frac{M_\phi (\rho_M^L-\rho_M^S)}{\rho^{n+1}} \xi_\phi^{n+1},
\end{eqnarray}
where $\xi_\phi^{n+1}$ is obtained from its definition in Eq.(\ref{Eq Phase change}). 
On the continuous level, as discussed in Section \ref{Sec Mass}, with $\rho$ and $\mathbf{m}_{\rho}$ from Eq.(\ref{Eq Material property}) and $\nabla \cdot \mathbf{u}$ in Eq.(\ref{Eq Divergence}), Eq.(\ref{Eq Mass}) is implied. However, this is not necessarily true after discretization when the phase change happens and the densities of the liquid and solid phases are not the same. As a result, the \textit{consistency of mass conservation} and \textit{consistency of mass and momentum transport} are violated. In order to remedy this issue, a momentum source $S_m \mathbf{u}^{*,n+1}$ is added to the momentum equation, where $S_m$ is the residual of the fully-discretized mass conservation equation, i.e., 
\begin{equation}\label{Eq Mass discrete}
S_m=\frac{\gamma_t \rho^{n+1}-\hat{\rho}}{\Delta t}+\tilde{\nabla} \cdot \tilde{\mathbf{m}}_{\rho}.
\end{equation}
It should be noted that $S_m$ only appears on the discrete level due to discretization errors, and it is exactly zero away from the liquid-solid interface.

Following the above steps, the physical connections among different parts of the proposed model are correctly captured at the discrete level. With similar analyses to those in the proofs of Theorem \ref{Theorem Two-Phase}, Theorem \ref{Theorem FSI}, and Theorem \ref{Theorem Solidification} in Section \ref{Sec Properties}, one can easily show that those theorems remain intact on the discrete level. This will be numerically verified in Section \ref{Sec Validation}.

\section{Results}\label{Sec Results}
In this section, various numerical tests are performed to verify and demonstrate the proposed model in Section \ref{Sec The proposed model}. Then, the predictions from the proposed model are compared to experimental data and other simulations. Finally, two challenging setups are performed to illustrate the capability of the proposed model. The formal order of accuracy and the conservation property of the discrete operators in Section \ref{Sec Discretizations} have been carefully verified in \citep{Huangetal2020,Huangetal2020CAC,Huangetal2020N} and therefore those verifications are not repeated here. Unless otherwise specified, the initial velocity is zero, and $\eta_\varphi=\eta_\phi=h$ and $M_\varphi \lambda_\varphi=10^{-9}$ are set, where $h$ denotes the grid/cell size.

All the upcoming numerical tests initialize the order parameters with a hyperbolic tangent profile, i.e., $\frac{1}{2} \left(1+ \tanh\left(\frac{\mathcal{D}}{\sqrt{2}\eta}\right) \right)$. Here, $\mathcal{D}$ is the signed distance function of ``$G$-$M$'' interface with $\eta=\eta_\varphi$ when initializing $\varphi$, while it is the signed distance function of the liquid-solid interface with $\eta=\eta_\phi$ to initialize $\phi$. When the liquid phase is initially absent, $\phi|_{t=0}=0$ is set, while $\phi|_{t=0}=1$ is used when there is no solid phase at the beginning. Similarly, if the gas phase is initially absent, we have $\varphi|_{t=0}=1$. The signed distance function in the present study is determined analytically, for example, it is $\mathcal{D}_c=r_c-\sqrt{(x-x_c)^2+(y-y_c)^2}$ for a circle centered at $(x_c,y_c)$ with a radius of $r_c$, and it is $\mathcal{D}_h=y_h-y$ for a horizontal line at $y_h$. Further, $\mathcal{D}=\min(\mathcal{D}_h,-\mathcal{D}_c)$ is the signed distance function positive below the horizontal line but outside the circle.

\subsection{Verification}\label{Sec Validation}
We first verify the theorems in Section \ref{Sec Properties} with problems having analytical solutions. Then, the correspondence of the mass conservation and the volume change of the phase change material ``$M$'' is illustrated. Finally, the effectiveness of the proposed surface tension and drag forces is demonstrated.

\subsubsection{Large-Density-Ratio advection}\label{Sec Advection}
We consider a large-density-ratio advection problem to verify Theorem \ref{Theorem Two-Phase} where the solid phase is absent and the temperature is above the melting temperature. 
The unit domain considered is doubly periodic. A circular drop, whose density is $\rho_M^L=10^4$, is initially at the center of the domain with a radius $0.2$, surrounded by a gas whose density is $\rho_G=1$. The specific heats of the phases are $1 \times 10^3$, and the material properties of the liquid phase are shared with the solid phase. The viscosity, heat conduction, surface tension, and gravity are neglected. Other parameters are $\Gamma_\phi=2.41 \times 10^{-7}$, $\mu_\phi=2.6 \times 10^{-5}$, $T_M=2$, $L=100$, and $\eta_\varphi=\eta_\phi=3h$. 
The solid phase is initially absent, i.e., $\alpha_S=\varphi-\varphi\phi=0$, and therefore we have initial $\phi$ being $1$. The initial velocity and temperature are $\mathbf{u}_0=\{u_0,v_0\}=\{1,1\}$ and $T_0=3>T_M$, respectively. 
The domain is discretized by $128 \times 128$ grid cells, and the time step is determined from $u_0\Delta t/h=0.1$. 

\begin{figure}[!t]
	\centering
	\includegraphics[scale=.5]{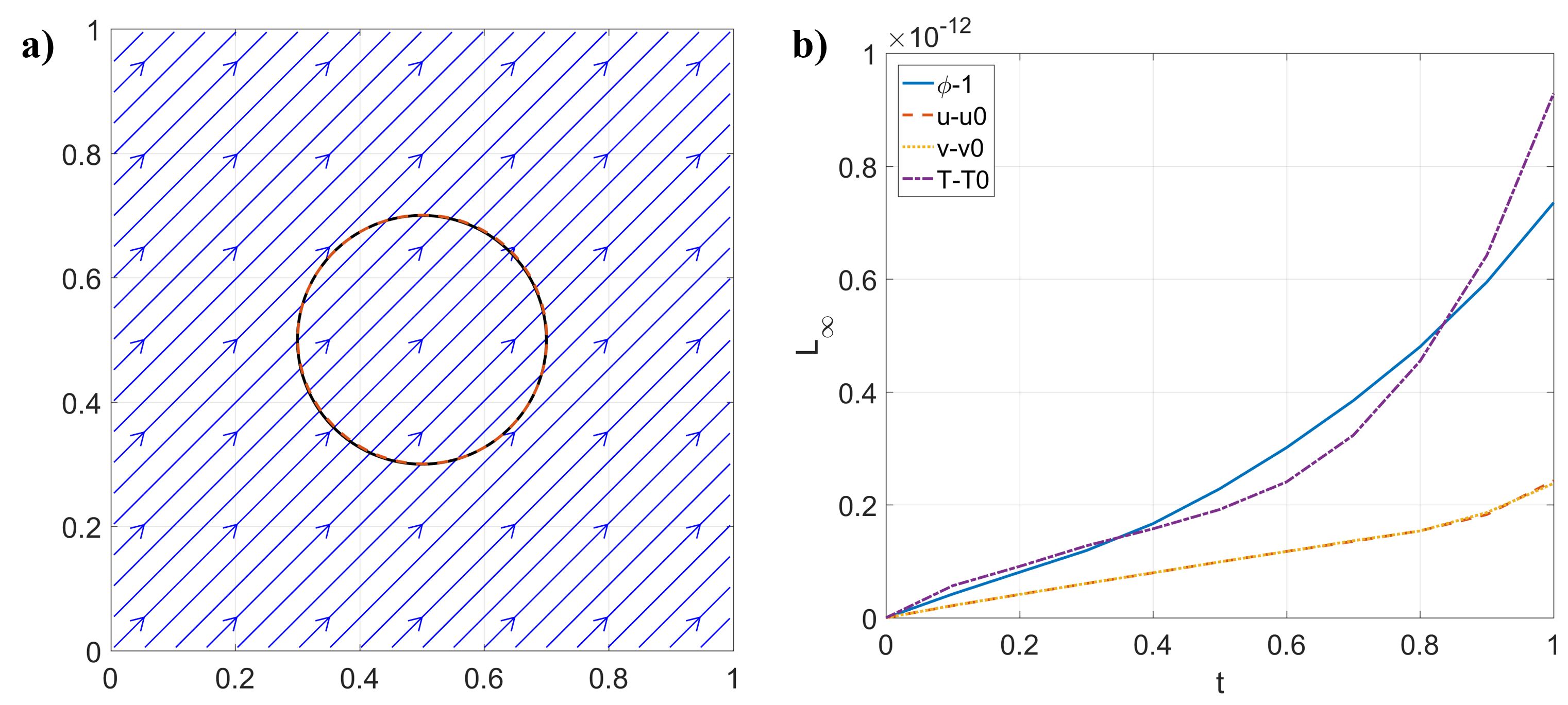}
	\caption{Results of the large-density-ratio advection. a) Streamlines at $t=1$ and interface of the drop ($\varphi=0.5$) at $t=0$ and $t=1$. Blue arrow lines: streamlines at $t=1$, Black solid line: interface at $t=0$, Red dashed line: interface at $t=1$. The drop returns to its original location at $t=1$ without deformation. b) $L_\infty$ norms of $(\phi-1)$, $(u-u_0)$, $(v-v_0)$, and $(T-T_0)$ versus time. The velocity preserves its initial value, the solid phase remains absent, and the temperate equilibrium is maintained exactly, which verify Theorem~\ref{Theorem Two-Phase}. \label{Fig Advection}}
\end{figure}

From Theorem \ref{Theorem Two-Phase}, the solid phase remains absent, i.e., $\alpha_S=\varphi-\varphi\phi=0$ or $\phi=1$ at $\forall t>0$, the temperature remains homogeneous, i.e., $T=T_0$ at $\forall t>0$, and the two-phase flow solution is produced, with the above setup. 
Expected results are obtained and shown in Fig.\ref{Fig Advection}. In this setup, the circular drop is translated by the homogeneous velocity. Therefore, there should not be any changes to the shape of the drop and the velocity. 
From Fig.\ref{Fig Advection} a), we observe that the drop correctly returns to its initial location at $t=1$, without any deformation, and the streamlines at $t=1$ remain straight and homogeneous. Quantitatively, the difference of the velocity from its initial value is the round-off error, as shown in Fig.\ref{Fig Advection} b). 
Fig.\ref{Fig Advection} b) also shows that Theorem \ref{Theorem Two-Phase} is true due to $\phi=1$ and $T=T_0$ at $\forall t>0$.

It is worth mentioning that the density ratio in this case is $10^4$, and there are no physical effects, e.g., viscosity and thermal conduction, to homogenize the solution. Without satisfying the consistency conditions, the drop will suffer from unphysical deformations, and the velocity and order parameter $\phi$ will become fluctuating, which are observed in \citep{Huangetal2020,Huangetal2020CAC,Huangetal2020N,Huangetal2020NPMC}. The same will happen to the temperature if Eq.(\ref{Eq Energy2}) is applied, instead of the proposed Eq.(\ref{Eq Energy}) that satisfies the consistency conditions, as analyzed in Section~\ref{Sec Energy}.

\subsubsection{Couette flow}\label{Sec Couette}
A Couette flow problem is solved to verify Theorem \ref{Theorem FSI} where the liquid phase is absent and the temperature is below the melting temperature. 
The unit domain considered is periodic along the $x$ axis while is no-slip along the $y$ axis. Both the top and bottom boundaries are adiabatic but the top one is moving with a unit horizontal velocity, i.e., $u_{top}=1$. The solid phase is at the bottom below $y=0.3$, while the gas phase fills the rest of the domain. The input parameters are $\rho=1$, $\mu=0.15$, $C_p=10^3$, $\kappa=0$, $\Gamma_\phi=2.41 \times 10^{-7}$, $\mu_\phi=2.6 \times 10^{-5}$, $T_M=2$, $L=100$, $\sigma=10^{-2}$, and $\mathbf{g}=\mathbf{0}$. 
The liquid phase is initially absent, i.e., $\alpha_L=\varphi \phi=0$, and therefore we have $\phi=0$. The initial temperature is $T_0=1$ which is lower than the melting temperature $T_M=2$. 
The domain is discretized by $128 \times 128$ grid cells, and the time step is determined by $u_{top} \Delta t/h=0.1$. 

\begin{figure}[!t]
	\centering
	\includegraphics[scale=.45]{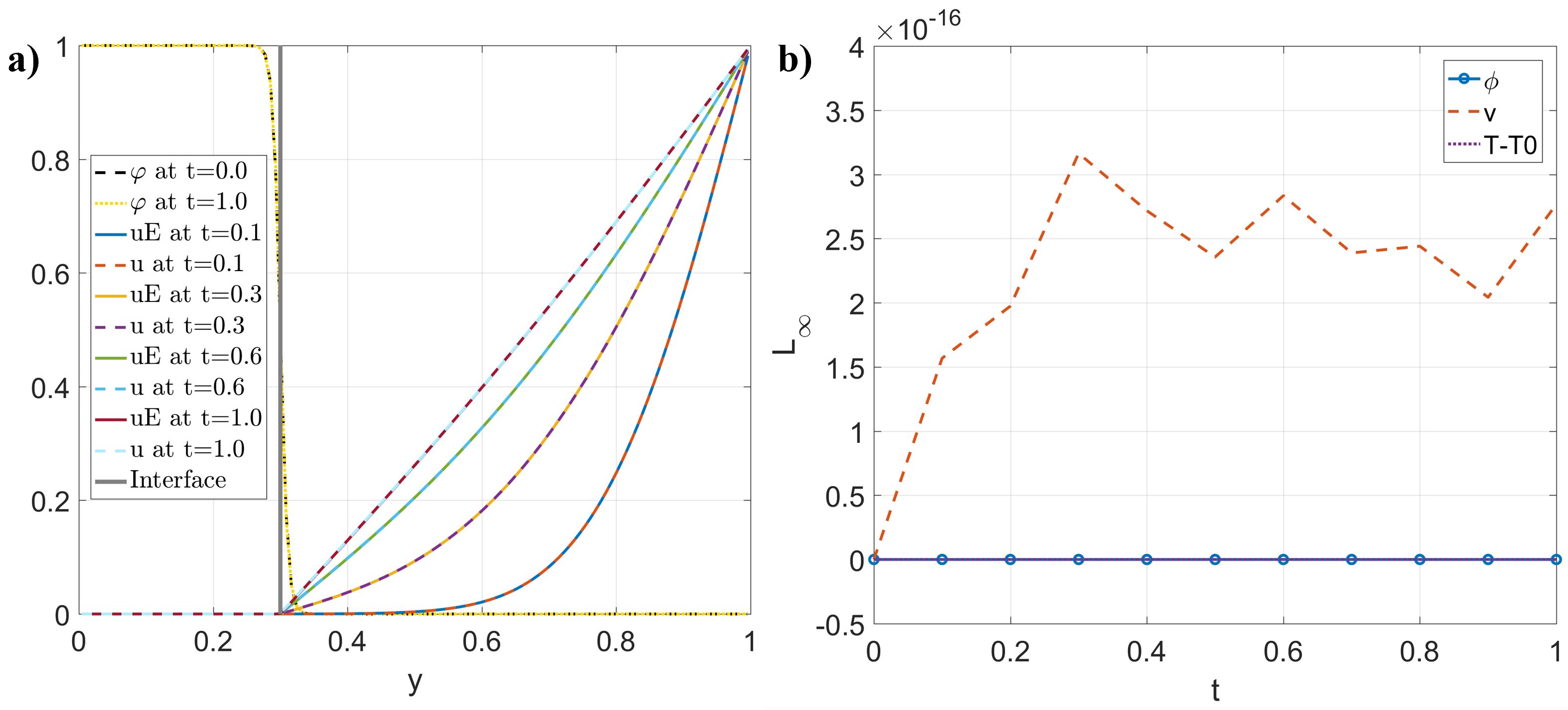}
	\caption{Results of the Couette flow. a) Profile of $\varphi$ at $t=0$ and $t=1$, and profiles of $u_E$ and $u$ at $t=0$, $t=0.1$, $t=0.3$, $t=0.6$, and $t=1$. Here, $u_E$ is the exact solution of the Couette flow in Eq.(\ref{Eq Couette solution}), and the numerical predictions and analytical solutions are overlapped. b) $L_\infty$ norms of $\phi$, $v$, and $(T-T_0)$ versus time. The liquid phase remains absent, the temperature equilibrium is preserved, and the unidirectional flow condition is produced, which verify Theorem \ref{Theorem FSI}.\label{Fig Couette}}
\end{figure}

The above setup is equivalent to the following Couette flow:
\begin{equation}\label{Eq Couette}
\frac{\partial u}{\partial t}=\nu \frac{\partial^2 u}{\partial y^2},
\quad
u=u_0 \quad \mathrm{at} \quad y=y_0,
\quad
u=u_1 \quad \mathrm{at} \quad y=y_1,
\quad
u|_{t=0}=0,
\quad
\nu=0.15,
\end{equation}
where $u_0=0$, $u_1=u_{top}=1$, $y_0=0.3$, and $y_1=1$.
The exact solution of Eq.(\ref{Eq Couette}) is 
\begin{equation}\label{Eq Couette solution}
u_E=\left\{ 
\begin{array}{cc}
\sum_{\alpha=1}^\infty \frac{2(u_1\cos(\alpha \pi)-u_0)}{\alpha \pi} \exp\left[-\nu \left(\frac{\alpha \pi}{y_1-y_0}\right)^2 t\right] \sin\left[\frac{\alpha \pi}{y_1-y_0} (y-y_0)\right]\\
+\frac{u_1-u_0}{y_1-y_0} (y-y_0) +u_0,  \quad y_0 \leqslant y \leqslant y_1,\\
0, \quad \mathrm{else},
\end{array} 
\right.
\end{equation}
derived from separation of variables. 
Theorem \ref{Theorem FSI} implies that the liquid phase remains absent, i.e., $\alpha_L=\varphi\phi=0$ or $\phi=0$ at $\forall t>0$, the temperature remains homogeneous, i.e., $T=T_0$ at $\forall t>0$, and the solution in Eq.(\ref{Eq Couette solution}) is produced (or approximated) by the FD/BP FSI formulation, with the present setup. 
Expected results are obtained and shown in Fig.\ref{Fig Couette}. First in Fig.\ref{Fig Couette} a), the profile of the solid fraction $(\alpha_S=\varphi-\varphi \phi=\varphi)$ at $t=1$ overlaps the one at $t=0$, representing that the solid phase is stationary. Moreover, the profiles of the solution from the proposed model are indistinguishable from the exact solution in Eq.(\ref{Eq Couette solution}), noticing that the summation in Eq.(\ref{Eq Couette solution}) is from $\alpha=1$ to $\alpha=10^6$. 
Theorem \ref{Theorem FSI} is true, as shown in Fig.\ref{Fig Couette} b) that $\phi=0$ and $T=T_0$ at $\forall t>0$. In addition, the unidirectional condition, which is required to obtain Eq.(\ref{Eq Couette}), is also demonstrated, as shown in Fig.\ref{Fig Couette} b) that $v=0$ at $\forall t>0$.

\subsubsection{Stefan problem}\label{Sec Stefan}
The Stefan problem is performed to verify Theorem \ref{Theorem Solidification} where both the gas phase and the flow are absent and the liquid and solid phases have matched material properties (except the thermal conductivities). The setup in \citep{Javierreetal2006} is followed. The unit domain is periodic along the $x$ direction. Both the top and bottom boundaries are free-slip and adiabatic. The material properties are: $\rho=1$, $\mu=1$, $C_p=1$, $\kappa_M^L=0.05$, $\kappa_M^S=1$, $T_M=1$, $L=0.53$, and $\mathbf{g}=\mathbf{0}$. Other parameters are $\eta_\phi=\sqrt{2a_\phi}\zeta_\phi$, $M_\phi=1/(\nu_\phi \zeta_\phi^2 L)$, and $\lambda_\phi=1/(\nu_\phi M_\phi)$, where $a_\phi=0.0625$, $\nu_\phi=1$, and $\zeta_\phi=0.002$, the same as those in \citep{Javierreetal2006}. Initially, the liquid-solid interface is at $y=s_0=0.2$, above which there is the solid phase having a temperature $T_0^S=1.1$, while below which there is the liquid phase having a temperature $T_0^L=1.53$. The domain is discretized by $5 \times 2000$ grid cells, and the time step is $\Delta t=5 \times 10^{-6}$.

\begin{figure}[!t]
	\centering
	\includegraphics[scale=.45]{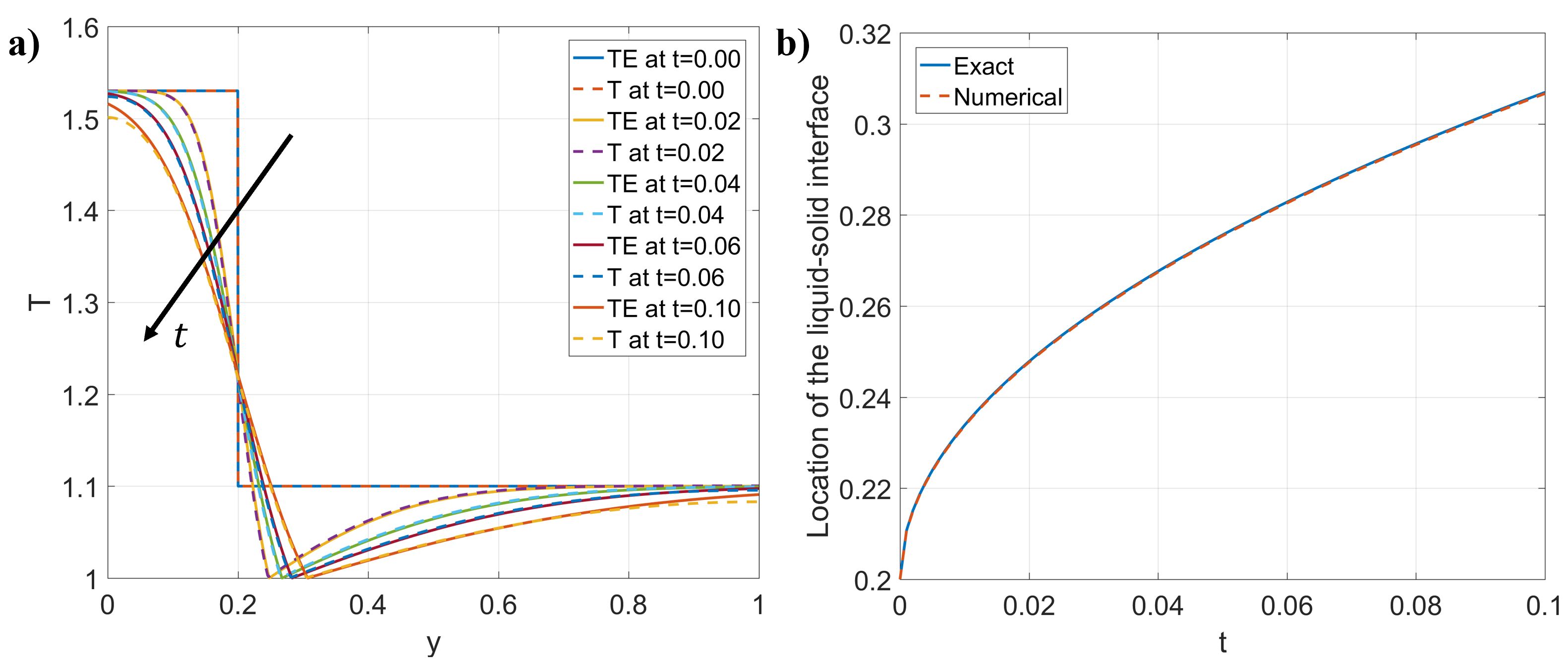}
	\caption{Results of the Stefan problem. a) Profiles of the temperature at different time. Here, $T_E$ is the exact solution of the Stefan problem in Eq.(\ref{Eq Stefan solution}). b) Location of the liquid-solid interface versus time. The numerical prediction agrees well with the analytical solution, which verifies Theorem \ref{Theorem Solidification}. \label{Fig Stefan}}
\end{figure}

The above setup is to produce the following Stefan problem \citep{Javierreetal2006} (based on $\Delta T=T-T_M$):
\begin{eqnarray}\label{Eq Stefan}
\frac{\partial \Delta T}{\partial t}=\frac{\partial}{\partial y} \left(\kappa_M^L \frac{\partial \Delta T}{\partial y}\right)
\quad y \in (-\infty,s(t)),
\quad
\frac{\partial \Delta T}{\partial t}=\frac{\partial}{\partial y} \left(\kappa_M^S \frac{\partial \Delta T}{\partial y}\right)
\quad y \in (s(t),+\infty),\\
\nonumber
\Delta T=T_0^L-T_M \quad \mathrm{at} \quad y \rightarrow -\infty,
\quad
\Delta T=T_0^S-T_M \quad \mathrm{at} \quad y \rightarrow +\infty,\\
\nonumber
\Delta T=0,
\quad
\mathrm{at} \quad y=s(t),\\
\nonumber
L\frac{ds}{dt}=\kappa_M^S \frac{\partial \Delta T}{\partial y}|_{y=s^+}-\kappa_M^L \frac{\partial \Delta T}{\partial y}|_{y=s^-},
\end{eqnarray}
where $s(t)$ is the location of the liquid-solid interface. Eq.(\ref{Eq Stefan}) has an analytical self-similar solution \citep{James1987,Javierreetal2006}:
\begin{eqnarray}\label{Eq Stefan solution}
s_E(t)=s_0+2\alpha\sqrt{t},\\
\nonumber
T_E-T_M=\left\{
\begin{array}{cc}
(T_0^L-T_M) \left[\frac{\mathrm{erfc} \left( \frac{y-s_0}{2\sqrt{\kappa_M^L t}} \right)-\mathrm{erfc} \left( \frac{\alpha}{\sqrt{\kappa_M^L}} \right) }{2-\mathrm{erfc} \left( \frac{\alpha}{\sqrt{\kappa_M^L}} \right)} \right]&  y<s(t),\\
(T_0^S-T_M)\left[1-\frac{\mathrm{erfc}\left( \frac{y-s_0}{2 \sqrt{\kappa_M^S t}} \right)}{\mathrm{erfc}\left(\frac{\alpha}{\sqrt{\kappa_M^S}}\right)} \right]    &  y>s(t),
\end{array}
\right.\\
\nonumber
\alpha=\frac{\sqrt{\kappa_M^S}}{\sqrt{\pi} L} \frac{T_M^S-T_M}{\mathrm{erfc}( \alpha/\sqrt{\kappa_M^S} )} \exp\left( -\frac{\alpha^2}{\kappa_M^S} \right)
+\frac{\sqrt{\kappa_M^L}}{\sqrt{\pi} L} \frac{T_M^L-T_M}{2-\mathrm{erfc}\left(\alpha/\sqrt{\kappa_M^L}\right)} \exp\left( -\frac{\alpha^2}{\kappa_M^L} \right).
\end{eqnarray}
Numerical results are compared to the exact solution Eq.(\ref{Eq Stefan solution}), and shown in Fig.\ref{Fig Stefan}. The interface location is specified as the $0.5$ contour of $(\varphi\phi)$ from the numerical results. Both the temperature and interface location agree with the exact solution very well. Minor discrepancy is observed near the domain boundary at $t=0.1$ in Fig.\ref{Fig Stefan} a) because in practice the domain is not infinite. Moreover, the $L_\infty$ norms of $(\varphi-1)$, $u$, and $v$ are on the orders of $10^{-16}$, $10^{-33}$, and $10^{-34}$, respectively, at the end of the simulation, which demonstrates Theorem \ref{Theorem Solidification}.

\subsubsection{The mass conservation and volume change}\label{Sec Mass conservation}
Here, we consider the effect of the mass conservation, which leads to the non-divergence-free velocity, i.e., Eq.(\ref{Eq Divergence}), when the phase change happens and the liquid and solid phases have different densities. Such an effect has been overlooked by many existing models. 
The unit domain considered is periodic along the $x$ axis. The bottom boundary is no-slip and has a fixed temperature $T_{bottom}=5$, while the top one is an outflow boundary having a fixed pressure $P_{top}=0$ and zero heat flux. 
At the bottom of the domain below $y=0.3$ is the liquid phase, whose material properties are $\rho_M^L=1000$, $\mu_M^L=1\times 10^{-3}$, $(C_p)_M^L=1$, and $\kappa_M^L=200$. Floating on the liquid phase is the solid phase, whose material properties are $\rho_M^S=900$, $\mu_M^S=1 \times 10^{-3}$, $(C_p)_M^S=1$, and $\kappa_M^S=300$. Above $y=0.6$ is the gas phase, whose material properties are $\rho_G=1$, $\mu_G=2 \times 10^{-5}$, $(C_p)_G=0.1$, and $\kappa_G=100$. Other parameters are $T_M=1$, $L=20$, $\sigma=0.0728$, $\rho_M^LL\Gamma_\phi/T_M=10$, $M_\phi=5 \times 10^{-4}$, and $\mathbf{g}=(0, -9.8)$. The initial temperature is $T_0=0.5$. The domain is discretized by $128 \times 128$ grid cells, and the time step is $\Delta t=10^{-3}$. 

\begin{figure}[!t]
	\centering
	\includegraphics[scale=.4]{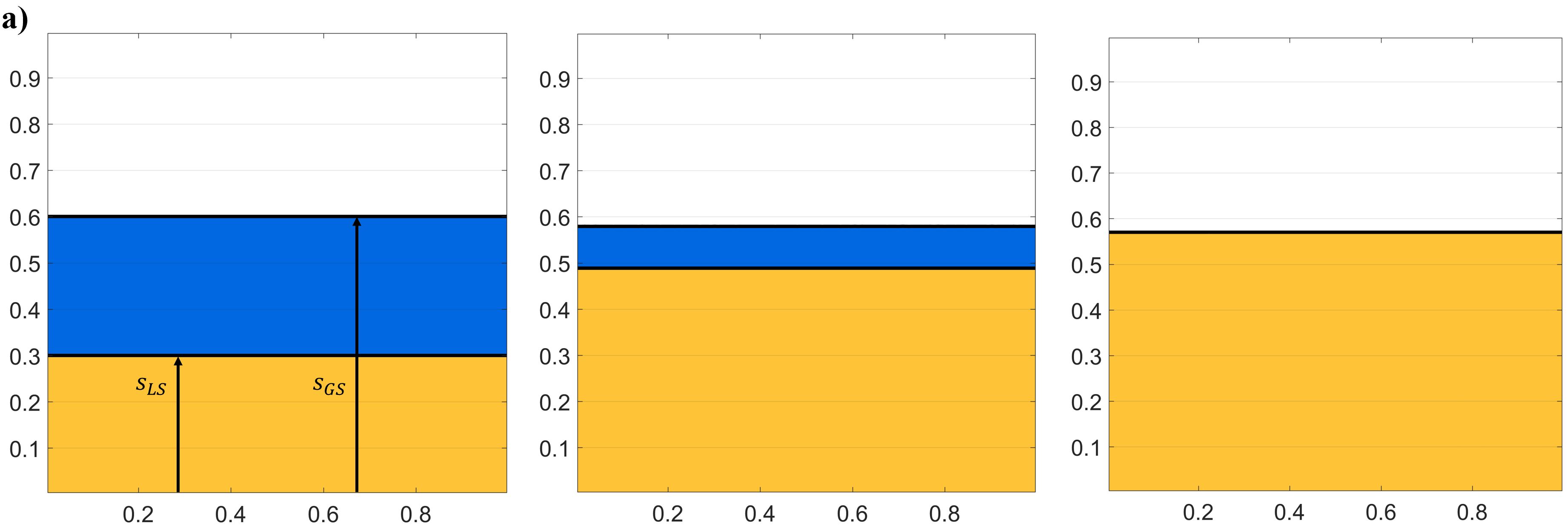}\\
	\includegraphics[scale=.4]{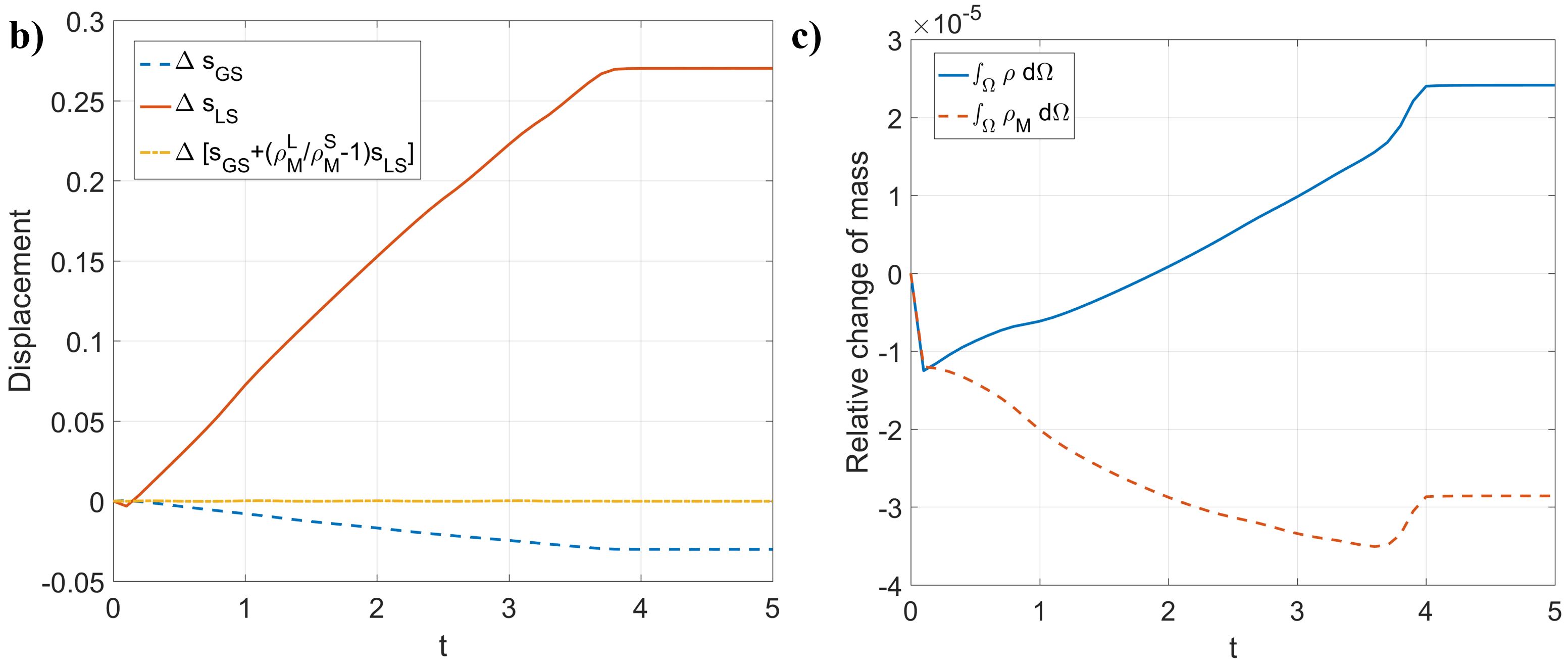}
	\caption{Results of the mass conservation and volume change. a) Snapshots of the phases at $t=0.0$, $t=2.5$, and $t=5.0$, and schematic of the locations of the liquid-solid ($s_{LS}$) and gas-solid ($s_{GS}$) interfaces. White: the gas phase, Orange: the liquid phase, Blue: the solid phase. b) Displacements of the gas-solid and liquid-solid interfaces versus time. The displacements of the interfaces are constrained by the mass conservation, quantified by $\Delta[s_{GS}+(\rho_M^L/\rho_M^S-1)s_{LS}]=0$. c) Relative changes of the total mass and the mass of the phase change material ``$M$'' versus time. The total mass is changing because the gas moves in and out of the domain following the volume change of the phase change material. The mass of the phase change material changes slightly (less than $0.004$\%) due to numerical errors. \label{Fig Mass}}
\end{figure}

The materials are heated by the bottom wall whose temperature is higher than the melting temperature. As a result, the solid phase will melt and finally disappear. Since the liquid density is 10\% larger than the solid phase, the final volume of the phase change material should be smaller than its initial value, in order to honor the mass conservation. 
Results are shown in Fig.\ref{Fig Mass} and match the expectation. 
From Fig.\ref{Fig Mass} a), the liquid-solid interface is moving upward while at the same time the gas-solid interface is moving downward. At the beginning, the volume (area) of ``$M$'', including its liquid and solid phases, is $0.6$. At the end of the simulation, there is only the liquid phase of ``$M$'' in the domain, and the gas-liquid interface stays horizontally below $y=0.6$, indicating that the volume of ``$M$'' is smaller than its initial value. 

Quantitative data are reported in Fig.\ref{Fig Mass} b) where the displacements of the liquid-solid and gas-solid interfaces versus time are plotted. The gas-solid interface is defined as the $0.5$ contour of $\varphi$, while the liquid-solid interface is the $0.5$ contour of $(\varphi\phi)$. We observe that the liquid-solid interface actually moves downward at the very beginning because the initial temperature is below the melting temperature. As a result, solidification happens in that period. As the materials are heated from the bottom wall, the solid melts, leading to the rise of the liquid-solid interface but fall of the gas-solid interface, as expected. The melting process ends before $t=4$. We estimate the mass of the phase change material ``$M$''  simply by $[\rho_M^L s_{LS}+\rho_M^S(s_{GS}-s_{LS})]$, where $s_{LS}$ and $s_{GS}$ denote the locations of the liquid-solid and gas-solid interfaces, respectively, illustrated in the first snapshot in Fig.\ref{Fig Mass} a). As plotted in Fig.\ref{Fig Mass} b), the change of $[s_{GS}+(\rho_M^L/\rho_M^S-1)s_{LS}]$ is negligible, which implies that the movements of the interfaces are constrained by the mass conservation. One can expect a more obvious displacement of the gas-solid interface, induced by the phase change, to appear if the density difference of the liquid and solid phases of the phase change material is larger than the present setup.

Although the proposed model strictly satisfies the mass conservation, i.e., Eq.(\ref{Eq Mass}), the present scheme does not always do, as discussed in Section \ref{Sec Discretizations}. Fig.\ref{Fig Mass} c) shows the relative changes of the total mass ($\int_{\Omega} \rho d\Omega$) and the mass of ``$M$'' ($\int_{\Omega} \rho_M d\Omega$) versus time. Here, $\rho_M=\rho_M^L (\varphi\phi)+\rho_M^S(\varphi-\varphi\phi)$ and the integral is computed from the mid-point rule. It should be noted that the total mass in this case is not conserved, and its change is related to the volume change of ``$M$'' during the phase change. The initial decrease of the total mass corresponds to the solidification process, where the volume of ``$M$'' expands and therefore the gas is squeezed out. As melting occurs, the total mass increases because the volume of ``$M$'' reduces, and the gas moves into the domain. When melting is completed, the total mass stops changing as well. On the other hand, the mass of ``$M$'' should be conserved even though the phase change happens, while it is not exactly true due to numerical errors. Nonetheless, its relative change is very small, on the order of $10^{-5}$, which is satisfactory. It should be noted that as long as the velocity is divergence-free, i.e., in the present work the phase change is absent or the densities of the liquid and solid phases are the same, the present scheme exactly conserves the mass of ``$M$'' as well as ``$G$'' on the discrete level, see \citep{Huangetal2020,Huangetal2020CAC}.

In summary, the proposed model automatically and physically captures the volume change induced by the phase change, and therefore the mass conservation, thanks to the \textit{consistency of mass conservation}. This physical behavior is not correctly captured in many existing models, where the velocity is assumed to be divergence-free. 

\subsubsection{Performances of surface tension and drag force models}\label{Sec Fources}
Here, we demonstrate the performances of the surface tension force $\mathbf{f}_s$, which models the surface tension at the gas-liquid interface, and the drag force $\mathbf{f}_d$, which enforces zero velocity in the solid phase and therefore the no-slip condition at the solid boundary, in the momentum equation Eq.(\ref{Eq Momentum}). Unless otherwise specified in this section, the following setup is employed. The material properties are $\rho_M^L=2.70 \times 10^3 \mathrm{kg/m^3}$, $\mu_M^L=1.4 \times 10^{-3} \mathrm{Pa \cdot s}$, $(C_p)_M^L=1.0424 \times 10^3 \mathrm{J/(K\cdot kg)}$, $\kappa_M^L=91 \mathrm{W/(m \cdot K)}$, $\rho_M^S=2.70 \times 10^3 \mathrm{kg/m^3}$, $\mu_M^S=1.4 \times 10^{-3} \mathrm{Pa \cdot s}$, $(C_p)_M^S=0.91 \times 10^3 \mathrm{J/(K\cdot kg)}$, $\kappa_M^S=211 \mathrm{W/(m \cdot K)}$, $\rho_G=0.4 \mathrm{kg/m^3}$, $\mu_G=4 \times 10^{-5} \mathrm{Pa \cdot s}$, $(C_p)_G=1.1 \times 10^3 \mathrm{J/(K\cdot kg)}$, $\kappa_G=61 \times 10^{-3} \mathrm{W/(m \cdot K)}$, $\Gamma_\phi=1.3 \times 10^{-3} \mathrm{m\cdot K}$, $\mu_\phi=1.3 \times 10^{-3} \mathrm{m/(s \cdot K)}$, $T_M=933.6 \mathrm{K}$, $L=3.8384 \times 10^5 \mathrm{J/kg}$, $\sigma=0.87 \mathrm{N/m}$, and $\mathbf{g}=(0, -9.8)$. The governing equations are non-dimensionalized by a density scale $1 \mathrm{kg/m^3}$, a length scale $0.01\mathrm{m}$, an acceleration scale $1\mathrm{m/s^2}$, and a temperature scale $933.6\mathrm{K}$. The domain considered is $[1 \times 1]$, and all the boundaries are no-slip and adiabatic except that the bottom one has a fixed temperature $T_{bottom}=0.5$. The center of a circular bubble having a radius $0.125$ is at $(0.5,0.275)$. The solid phase is below $y=0.1$, while the liquid phase fills the rest of the domain. The initial temperature is $0.5$ inside the solid phase, while is $1.1$ elsewhere. Notice that the non-dimensinalized melting temperature is $1$. The domain is discretized by $128 \times 128$ grid cells, and the time step is $\Delta t=1 \times 10^{-4}$.

\begin{figure}[!t]
	\centering
	\includegraphics[scale=.4]{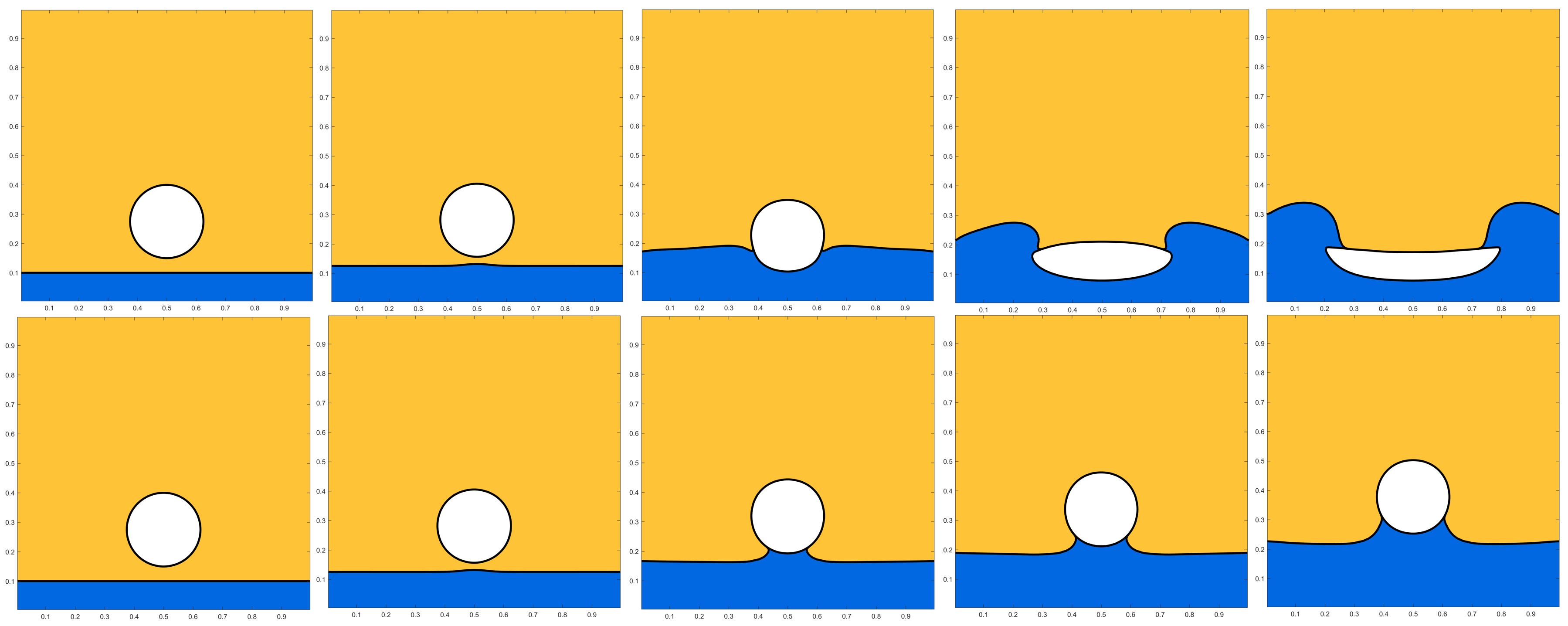}\\
	\caption{Results of cases 1 and 2 at $t=0.00$, $t=0.05$, $t=0.20$, $t=0.30$, and $t=0.50$. White: the gas phase, Orange: the liquid phase, Blue: the solid phase. Top: case 1 using the proposed surface tension force $\mathbf{f}_s=\phi \xi_\varphi \nabla \varphi$. Bottom: case 2 using the original surface tension force $\mathbf{f}_s=\xi_\varphi \nabla \varphi$. The original surface tension force (case 2) also appears at the gas-solid interface and therefore the bubble is less deformable than the one in case 1. The solid behaves like a fluid although its viscosity is about $1000$ times larger than the liquid phase. \label{Fig SurfaceForce}}
\end{figure}

\begin{figure}[!t]
	\centering
	\includegraphics[scale=.4]{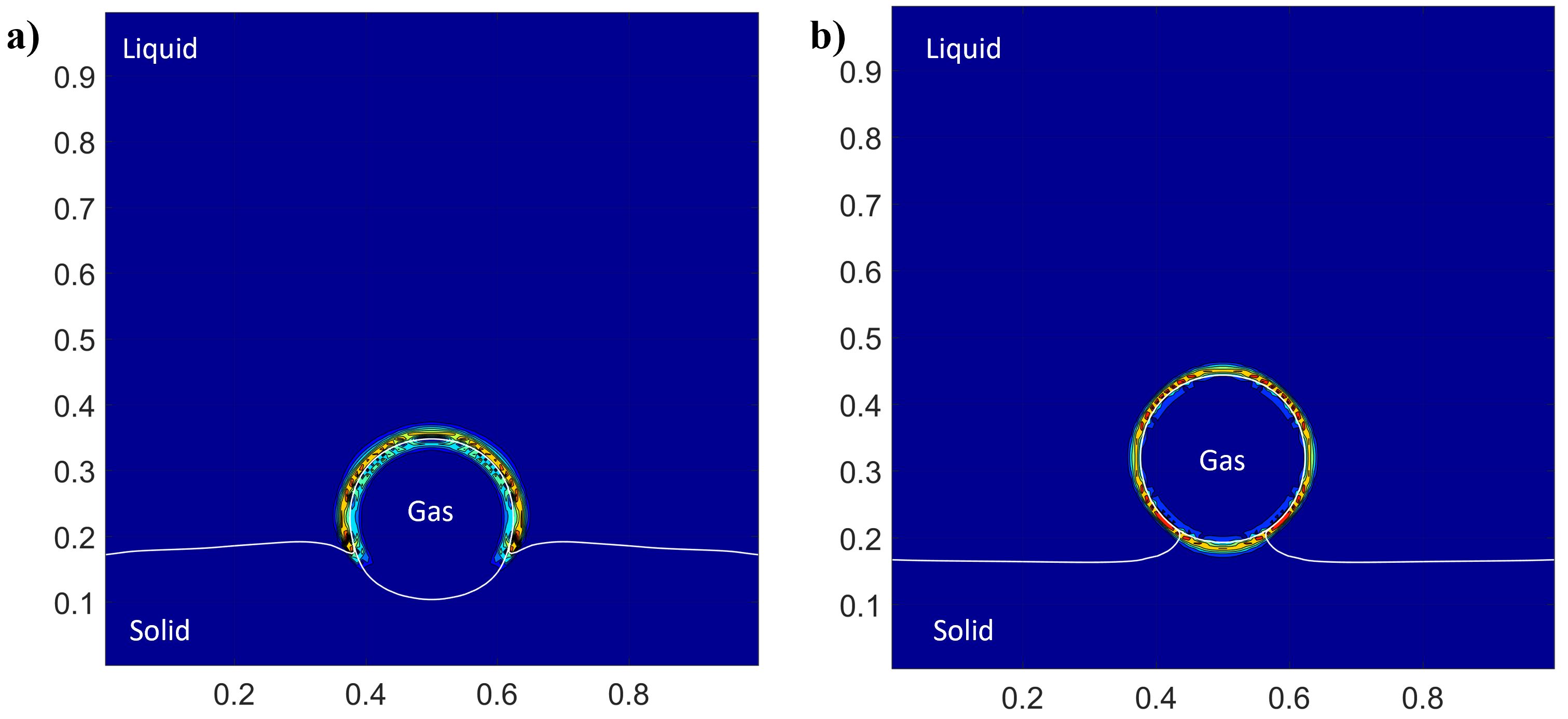}\\
	\caption{Magnitude of the surface tension forces $|\mathbf{f}_s|$ at $t=0.20$. a) Case 1: $\mathbf{f}_s=\phi \xi_\varphi \nabla \varphi$, b) Case 2: $\mathbf{f}_s=\xi_\varphi \nabla \varphi$. The proposed surface tension force in case 1 appears only at the gas-liquid interface, while the original one in case 2 mistakenly appears at the gas-solid interface.\label{Fig Fs2}}
\end{figure}

In the first two cases, the drag force is zero, i.e., $\mathbf{f}_d=\mathbf{0}$, while the solid viscosity becomes $1 \mathrm{Pa \cdot s}$, which is about $1000$ times larger than the liquid phase. In case 1, we employ the proposed surface tension force $\mathbf{f}_s=\phi \xi_\varphi \nabla \varphi$ in Eq.(\ref{Eq Momentum}), while it is $\mathbf{f}_s=\xi_\varphi \nabla \varphi$ in case 2. 
Results are shown in Fig.\ref{Fig SurfaceForce}, and the difference between case 1 ($\mathbf{f}_s=\phi \xi_\varphi \nabla \varphi$) and case 2 ($\mathbf{f}_s=\xi_\varphi \nabla \varphi$) is obvious. In case 1, the surface tension force only acts at the upper part of the bubble, contacting the liquid phase, which is desirable. As a result, the bottom part of the bubble is easier to be deformed, while the upper part tends to be flattened. On the other hand in case 2, the surface tension force acts on the entire bubble interface, no matter whether the bubble is contacting the liquid or solid phase. Consequently, the bubble remains circular even after it is contacted by the solid phase at its bottom part. The above analysis is demonstrated in Fig.\ref{Fig Fs2}, where the magnitude of the surface tension forces, i.e., $|\mathbf{f}_s|$, at $t=0.20$ in cases 1 and 2 is shown. It can be learned from Fig.\ref{Fig SurfaceForce} that the surface tension force can be influential to the results, and Fig.\ref{Fig Fs2} demonstrates that the proposed surface tension force, i.e., $\mathbf{f}_s=\phi \xi_\varphi \nabla \varphi$ in case 1, is the one that should be chosen. An alternative formulation, i.e., $\mathbf{f}_s=\xi_\varphi \nabla (\varphi \phi)$, has also been tested, and it produced an unstable solution. 
Another issue, observed in Fig.\ref{Fig SurfaceForce}, is that the solid phase behaves like a fluid, even though it is about 1000 times more viscous than the liquid phase, and $2.5 \times 10^4$ times more than the gas phase. In the ideal situation, the viscous force is infinite inside the solid phase, which in turn enforces zero velocity gradient there. In other words, after discretization, $\tilde{\mu}_M^S/h^2>>\tilde{\rho}_M^S/\Delta t$ should be true, where $\tilde{\mu}$ and $\tilde{\rho}$ are the non-dimensionalized viscosity and density, and in this specific case, $\mu_M^S>>1.6479 \mathrm{Pa \cdot s}$. We again tried $\mu_M^S=1000 \mathrm{Pa \cdot s}$ and it quickly became unstable. Therefore, increasing the solid viscosity is not an effective way to enforce zero velocity in the solid phase. This is the reason the drag force $\mathbf{f}_d$ is introduced in the proposed model.

\begin{figure}[!t]
	\centering
	\includegraphics[scale=.4]{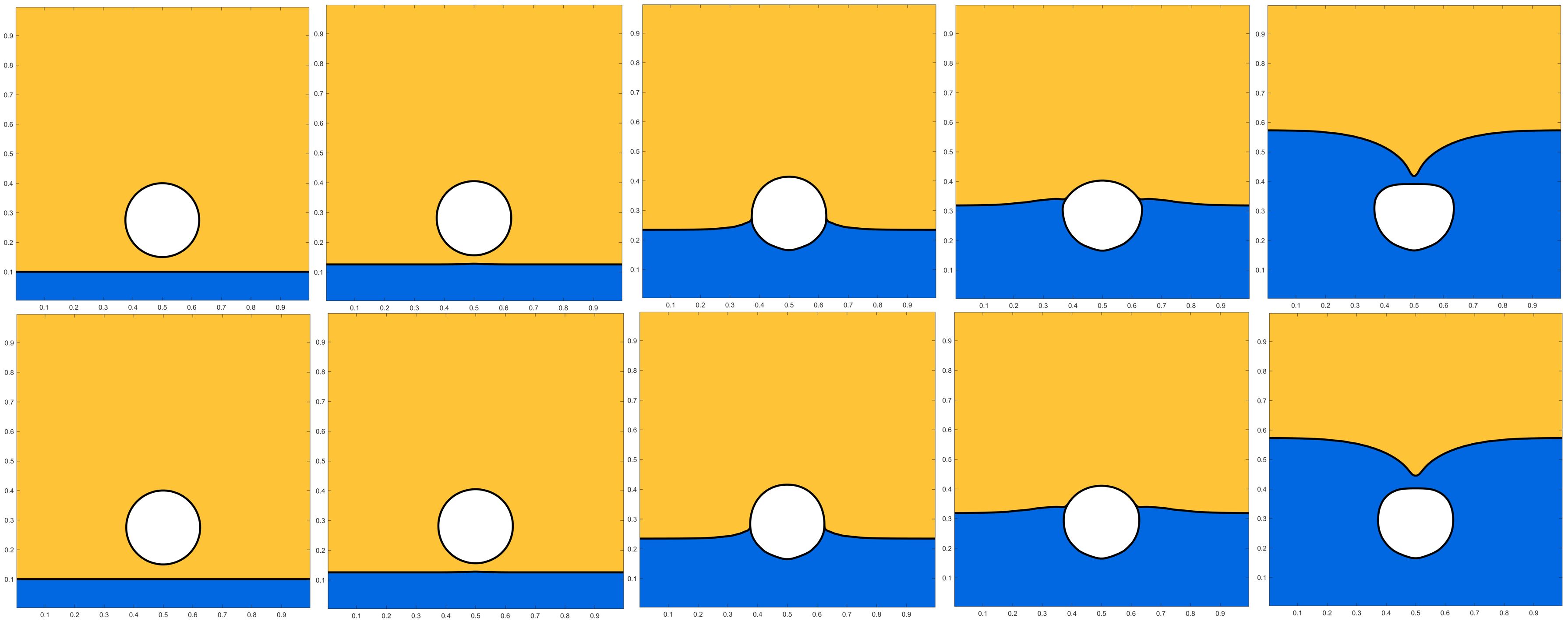}\\
	\caption{Results of cases 3 and 4 at $t=0.00$, $t=0.05$, $t=0.50$, $t=1.00$, and $t=3.40$. White: the gas phase, Orange: the liquid phase, Blue: the solid phase. Top: case 3 using $A_d = C_d \frac{\alpha_S^2}{(1-\alpha_S)^3+e_d}$. Bottom: case 4 using $A_d=\varphi C_d \frac{(1-\phi)^2}{\phi^3+e_d}$. Both choices of $A_d$ successfully suppress the solid motion and produce similar results, but the one in case 4 has a larger effective region resulting in a less-deformed bubble.\label{Fig DragForce}}
\end{figure}

Next, the drag force $\mathbf{f}_d$ is activated. In case 3, we apply the formulation of $A_d$ in Eq.(\ref{Eq Momentum}), while in case 4, an alternative definition of $A_d$, i.e., $A_d=\varphi C_d \frac{(1-\phi)^2}{\phi^3+e_d}$, is considered. The alternative $A_d$ can be easily derived from the drag force model in $\Omega_M$ proposed by Voller and Prakash \citep{VollerPrakash1987} and the diffuse domain approach \citep{Lietal2009}. Results are shown in Fig.\ref{Fig DragForce}. It is obvious that the solid movement is suppressed after comparing Fig.\ref{Fig DragForce} to Fig.\ref{Fig SurfaceForce}, and this has also been quantitatively demonstrated in Section \ref{Sec Couette}. As long as the bubble is ``caught'' by the solid from the bottom, it stops rising, unlike the one in Fig.\ref{Fig SurfaceForce}. Although both cases 3 and 4 produce similar results, one can observe that the bubble in case 4 is less deformed than the one in case 3 using the proposed formulation. This implies that the alternative $A_d$ in case 4 has a larger effective region to enforce the velocity to be zero, while its influence on the overall dynamics is negligible. We conclude that both choices of $A_d$ are valid, but we keep using the one in Eq.(\ref{Eq Momentum}) in the present study because it is equivalent to the Carman-Kozeny equation \citep{Carman1997}.

\subsection{Melting of gallium in a rectangular cavity}\label{Sec Gallium}
Here, we compare results from the proposed model to the experimental \citep{GauViskanta1986} and numerical \citep{Brentetal1988,Kimetal2011} results. Details of the setup have been given in \citep{GauViskanta1986,Brentetal1988}, and we follow those in the present study.
A rectangular cavity, whose width is $8.89\mathrm{cm}$ and height is $6.35\mathrm{cm}$, is initially filled with solid gallium whose temperature is $301.45\mathrm{K}$. The left wall has a fixed temperature $311.15\mathrm{K}$, it is $301.45\mathrm{K}$ at the right wall, while both the top and bottom walls are adiabatic. The material properties of gallium are: density $6093\mathrm{kg/m^3}$, viscosity $1.81 \times 10^{-3} \mathrm{Pa \cdot s}$, specific heat $381.5\mathrm{J/(K\cdot kg)}$, thermal conductivity $32\mathrm{W/(m \cdot K)}$, melting temperature $302.93\mathrm{K}$, and latent heat $80160\mathrm{J/kg}$. The gravity is $\mathbf{g}=(0, -9.8)\mathrm{kg/m^2}$, and the buoyancy force is computed from the Boussinesq approximation, i.e., $\mathbf{F}_g=-\rho_0 \beta (T-T_0) \mathbf{g}$, where $\rho_0=6095\mathrm{kg/m^3}$, $T_0=302.93\mathrm{K}$, and $\beta=1.2 \times 10^{-4} \mathrm{/K}$ is the thermal expansion coefficient. $\Gamma_\phi$ and $\mu_\phi$ are chosen to be $1.2 \times 10^{-4} \mathrm{m \cdot K}$ and $1.2 \times 10^{-4} \mathrm{m/(s \cdot K)}$.
The governing equations are non-dimensionalized by a density scale $6095\mathrm{kg/m^3}$, a length scale $0.01\mathrm{m}$, an acceleration scale $0.01 \mathrm{m/s^2}$, and a temperature scale $302.93 \mathrm{K}$. The domain is discretized by $126 \times 96$ grid cells. The initial time step is $\Delta t=0.05$ and adaptively changes to be $0.5h/\max(u,v)$. 

\begin{figure}[!t]
	\centering
	\includegraphics[scale=.4]{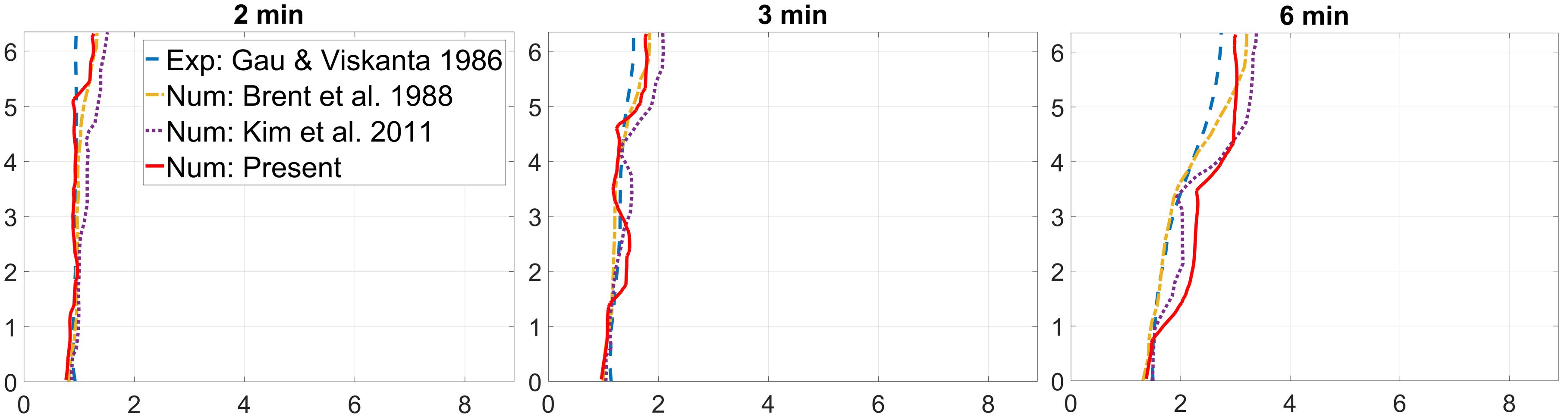}\\
	\includegraphics[scale=.4]{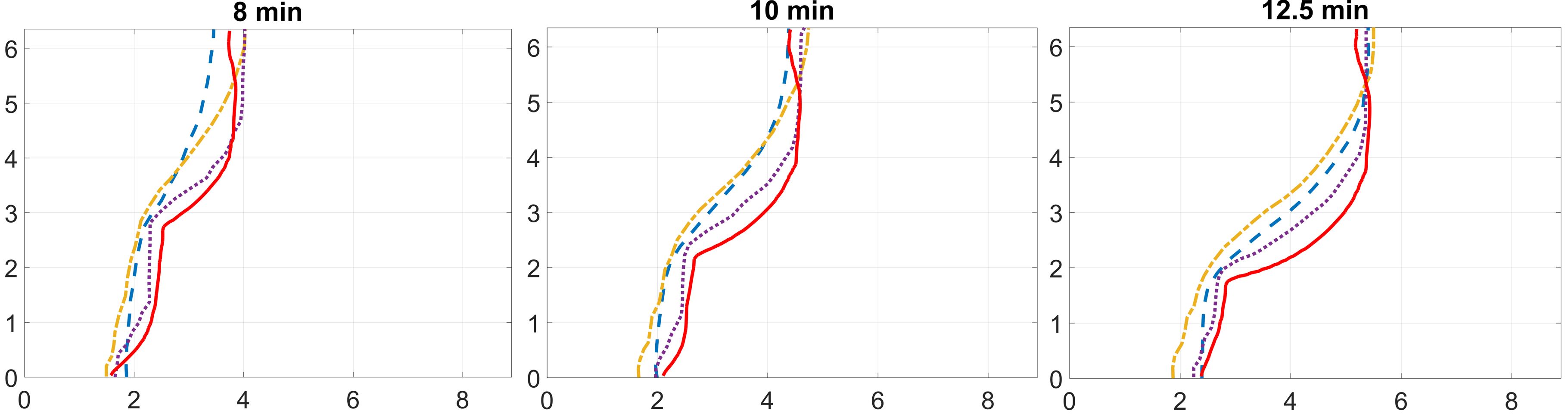}\\
	\includegraphics[scale=.4]{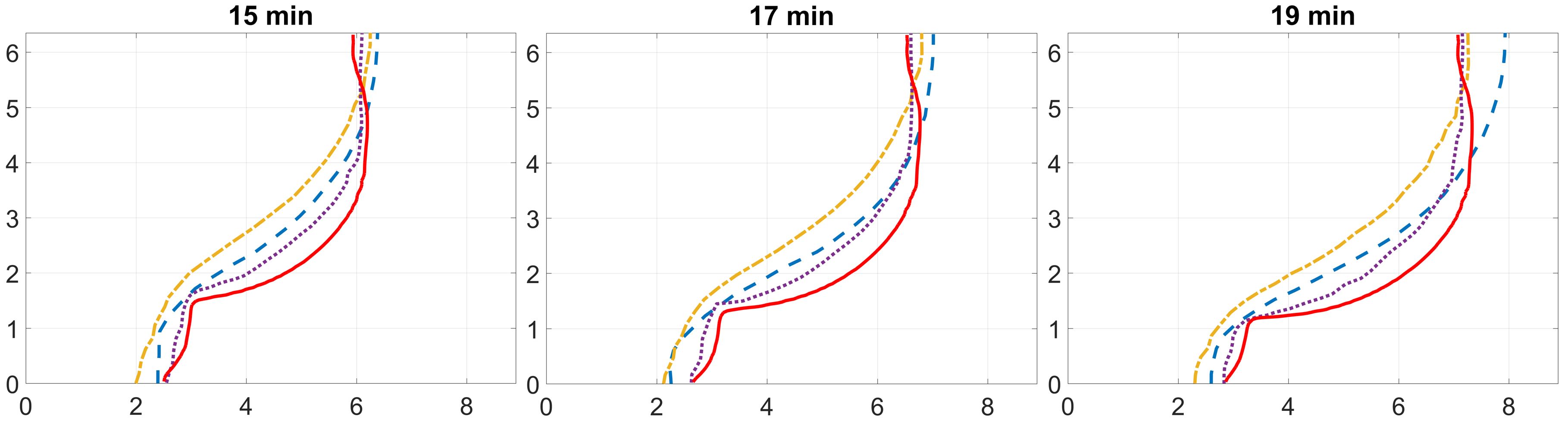}\\
	\caption{Results of melting of gallium in a rectangular cavity and the liquid-solid interface is presented. Both Brent et al. (1988) \citep{Brentetal1988} and Kim et al. (2011) \citep{Kimetal2011} use the enthalpy-porosity technique but Brent et al. (1988) \citep{Brentetal1988} neglect the convection of the liquid fraction in the energy equation. The present results behave similarly to those in Kim et al. (2011) \citep{Kimetal2011} and a stronger melt convection is predicted. Nevertheless, all the numerical predictions are consistent with the experimental data.\label{Fig Gallium}}
\end{figure}

Results are shown in Fig.\ref{Fig Gallium} where the liquid-solid interface at selected moments is presented, and a reasonable agreement is reached with the experimental and numerical data. Both \citep{Brentetal1988} and \citep{Kimetal2011} employed the enthalpy-porosity technique where the liquid fraction is algebraically determined by the local temperature. The results from \citep{Brentetal1988} is smoother but moves slower, probably attributed to neglecting the convection of the liquid fraction in the energy equation. The present results are close to those in \citep{Kimetal2011}. Both predict a similar melting speed but a more vertical interface than the experimental one. 

It is worth mentioning that the melting will not happen if the unmodified interpolation function $p(\phi)$ in Eq.(\ref{Eq Allen-Cahn}) is applied, because its derivative is zero in the solid-state. This demonstrates the significance of using the proposed $\tilde{p}'(\phi)$ in Eq.(\ref{Eq Phase change}) in realistic problems. The more detailed analysis has been provided in Section~\ref{Sec Phase change}.
It is preferable to understand $\mu_\phi$ and $\Gamma_\phi$ as tunable parameters of the proposed model, instead of their physical meaning, since the practical interface thickness is much larger than the physical value. In practice, we tune $\mu_\phi$ and $\Gamma_\phi$ so that the numerical result matches the experimental one at $t=2\mathrm{min}$, and obtain the rest of the results with those parameters. When tuning $\mu_\phi$ and $\Gamma_\phi$, we discover that $\mu_\phi$ controls the speed of the phase change, while $\Gamma_\phi$ affects the interface thickness. A larger $\mu_\phi$ gives a larger $M_\phi$, and as a result accelerates the phase change. We observe an over-compressed interface when $\Gamma_\phi$ is too small, while a too large $\Gamma_\phi$ casts difficulty to initialize the interface. This can be explained by the energy mechanism in the Phase-Field model of solidification Eq.(\ref{Eq Allen-Cahn}). $\lambda_\phi$ controls the net effect of the thermodynamical compression and diffusion that preserve the interface thickness. A too small $\lambda_\phi$, resulting from a small $\Gamma_\phi$, basically removes those compression and diffusion effects and leads to a sharp interface. On the other hand, a large $\lambda_\phi$ strengthens those effects, and a larger overheat is therefore needed to drive the order parameter, jumping from one equilibrium state to another across the double-well potential. We suggest $\mu_\phi$ and $\Gamma_\phi$ sharing the same value. We also test the effect of $e_d$ in the drag force $\mathbf{f}_d$, and little difference is observed when reducing $e_d$ from the default value $10^{-3}$ to $10^{-6}$.

\subsection{Rising bubbles with solidification}\label{Sec Bubbles}
Here, we consider a complicated case including interactions among the gas, liquid, and solid phases. The material properties of the liquid phase are $\rho_M^L=2.475 \times 10^3 \mathrm{kg/m^3}$, $\mu_M^L=1.4 \times 10^{-3} \mathrm{Pa \cdot s}$, $(C_p)_M^L=1.0424 \times 10^3 \mathrm{J/(K \cdot kg)}$, and $\kappa_M^L=91 \mathrm{W/(m \cdot K)}$. They are $\rho_M^S=2.70 \times 10^3 \mathrm{kg/m^3}$, $\mu_M^S=1.4 \times 10^{-3} \mathrm{Pa \cdot s}$, $(C_p)_M^S=0.91 \times 10^3 \mathrm{J/(K \cdot kg)}$, and $\kappa_M^S=211 \mathrm{W/(m \cdot K)}$ for the solid phase, and $\rho_G=0.4 \mathrm{kg/m^3}$, $\mu_G=4 \times 10^{-5} \mathrm{Pa \cdot s}$, $(C_p)_G=1.1 \times 10^3 \mathrm{J/(K \cdot kg)}$, and $\kappa_G=61 \times 10^{-3} \mathrm{W/(m \cdot K)}$ for the gas phase. The melting temperature is $T_M=933.6\mathrm{K}$, latent heat is $L=3.8384 \times 10^5 \mathrm{J/kg}$, the surface tension is $\sigma=0.87 \mathrm{N/m}$, the gravity is $\mathbf{g}=(0,-9.8)\mathrm{m/s^2}$, and the Gibbs-Thomson and linear kinetic coefficients are chosen to be $\Gamma_\phi=1.3 \times 10^{-3} \mathrm{m \cdot K}$ and $\mu_\phi=1.3 \times 10^{-3} \mathrm{m/(s \cdot K)}$, respectively. The governing equations are non-dimensionalized by a density scale $1\mathrm{kg/m^3}$, a length scale $0.01\mathrm{m}$, an acceleration scale $1\mathrm{m/s^2}$, and a temperature scale $933.6 \mathrm{K}$ the same as the melting temperature.

\begin{figure}[!t]
	\centering
	\includegraphics[scale=.43]{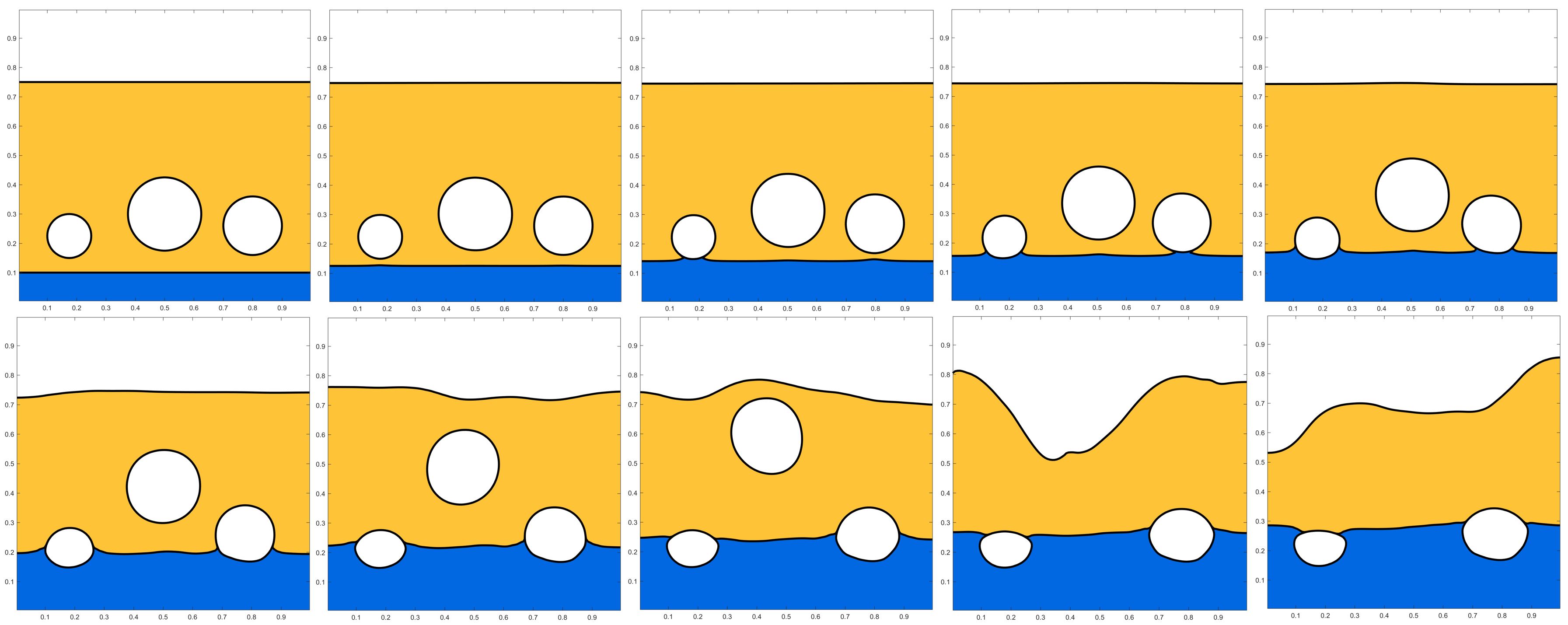}
	\includegraphics[scale=.43]{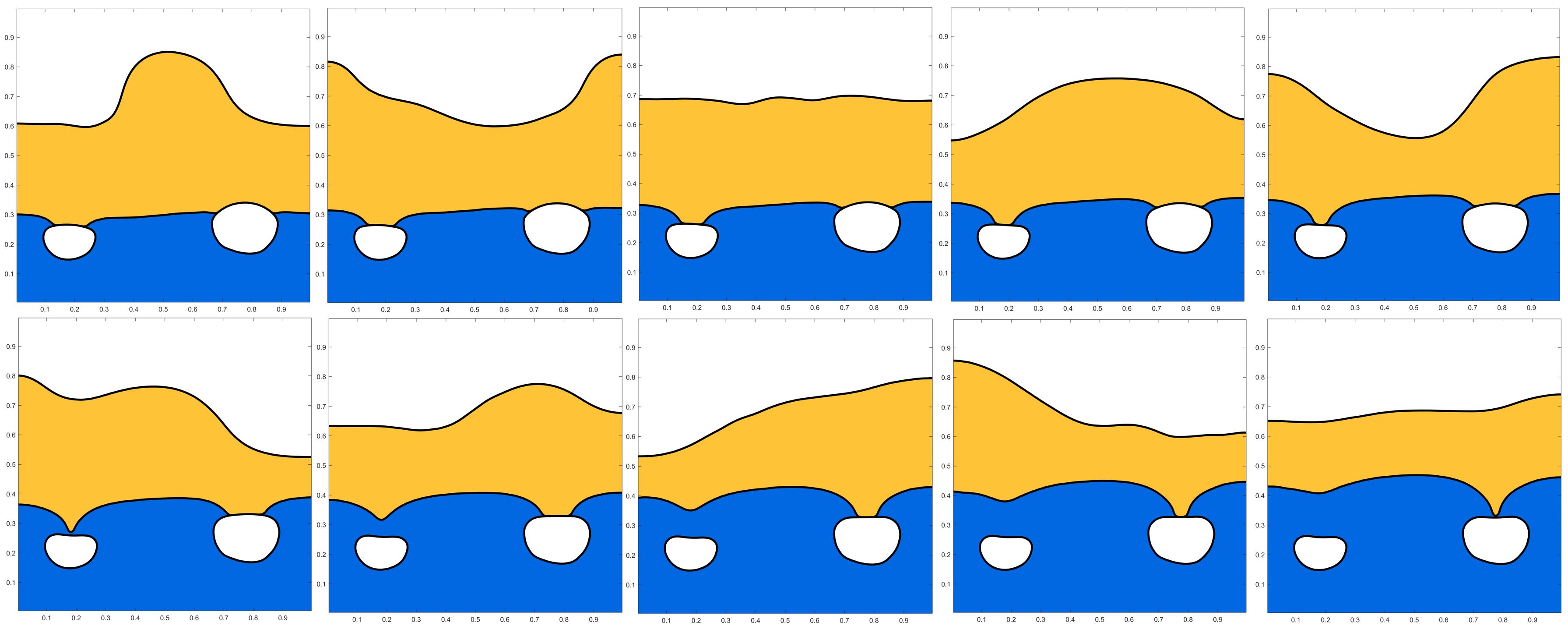}
	\includegraphics[scale=.43]{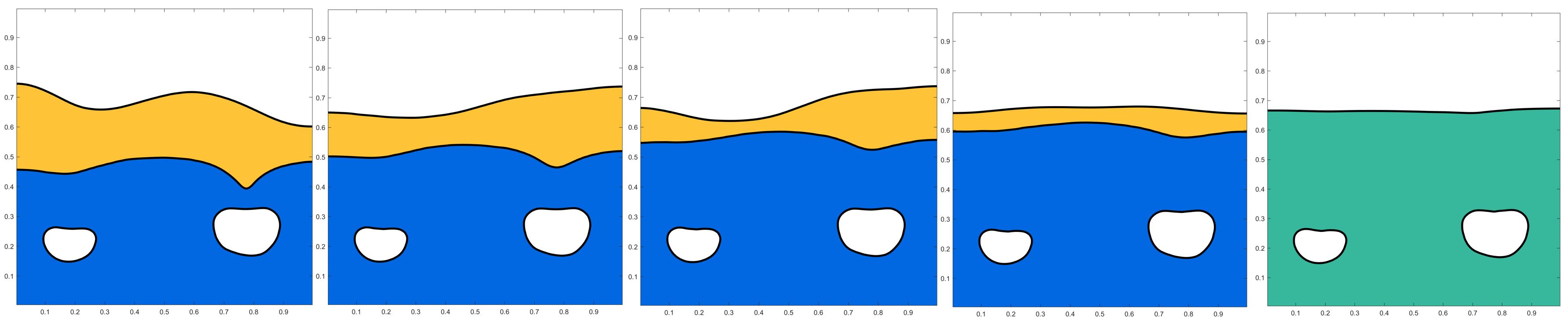}
	\caption{Results of rising bubbles with solidification. White: the gas phase, Orange: the liquid phase, Blue: the solid phase, Green: the solid phase when the phase change is finished. From left to right and top to bottom, $t=0.00$, $t=0.05$, $t=0.10$, $t=0.15$, $t=0.20$, $t=0.30$, $t=0.40$, $t=0.50$, $t=0.60$, $t=0.70$, $t=0.80$, $t=0.90$, $t=1.00$, $t=1.10$, $t=1.20$, $t=1.40$, $t=1.60$, $t=1.80$, $t=2.00$, $t=2.20$, $t=2.50$, $t=3.00$, $t=3.50$, $t=4.00$, and $t=5.00$. \label{Fig Bubbles}}
\end{figure}

A unit domain is considered. Both the left and right boundaries are no-slip and adiabatic walls. The bottom boundary is no-slip with a fixed temperature $T_{bottom}=0.5$. The top boundary has a fixed pressure $P_{top}=0$ and a zero heat flux. The domain is discretized by $128 \times 128$ cells, and the time step is $\Delta t=10^{-4}$. The initial condition of the phases is illustrated in the first snapshot of Fig.\ref{Fig Bubbles}. Above $y=0.75$ is the gas phase, while the solid phase is at the bottom below $y=0.1$. In the middle of the domain is the liquid phase inside which there are three circular gas bubbles. The radii of the bubbles from left to right are $0.075$, $0.125$, and $0.1$, and their centers are at $(0.175,0.225)$, $(0.5,0.3)$, and $(0.8,0.26)$, respectively. The initial temperature is 0.5, the same as $T_{bottom}$, inside the solid phase, while it is 1.1 elsewhere. Note that the non-dimensionalized melting temperature is $1$.

Results are shown in Fig.\ref{Fig Bubbles}. The gas, liquid, and solid phases are filled by the white, orange, and blue colors, and the solid phase becomes green when the phase change is finished. The bubbles are moving upward due to the buoyancy effect, and, at the same time, the liquid is solidifying as its temperature is cooled down by the bottom wall. The motion of the bubbles drives the liquid and produces melt convection. As a result, the liquid below the bubbles solidifies faster than its neighbor, and the liquid-solid interface first ``catches'' the left bubble then the right one. As the largest bubble at the middle rises, the gas-liquid interface above starts to be perturbed, which, in turn, deviates the bubble rising from its vertical line. When the middle bubble merges the gas-liquid interface, a strong capillary wave is produced due to the surface tension. As the capillary wave travels back and forth, the liquid-solid interface keeps moving upward. Since the heat conductivity of the gas is much smaller than the liquid or solid, the solidification is slower right above the two trapped gas bubbles, and the liquid-solid interface forms a ``V'' shape there. As the liquid-solid interface gets closer to the gas-liquid one, the capillary wave is quickly attenuated by the viscosity, due to the zero velocity of the solid. At the end of the simulation, the liquid completely solidifies with two hollows formed by the right and left bubbles. 

\subsection{Melting and solidification}
\begin{figure}[!t]
	\centering
	\includegraphics[scale=.43]{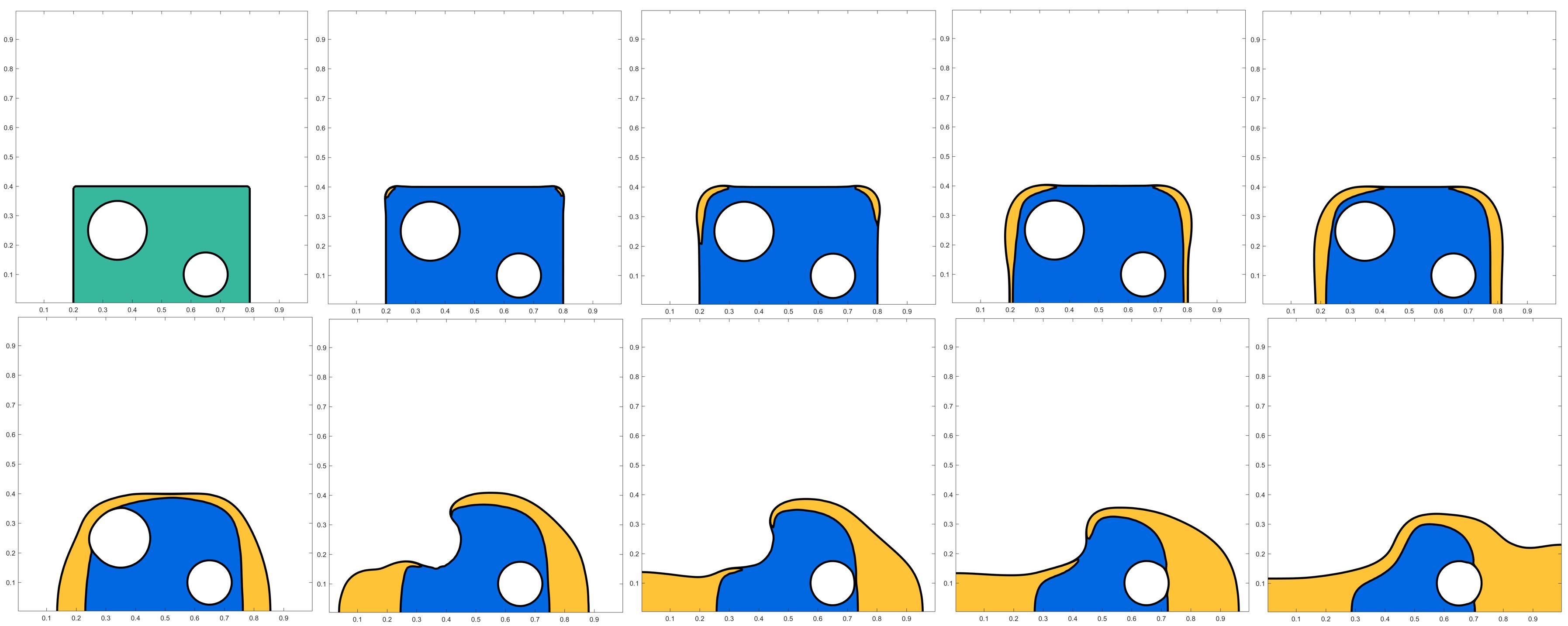}
	\includegraphics[scale=.43]{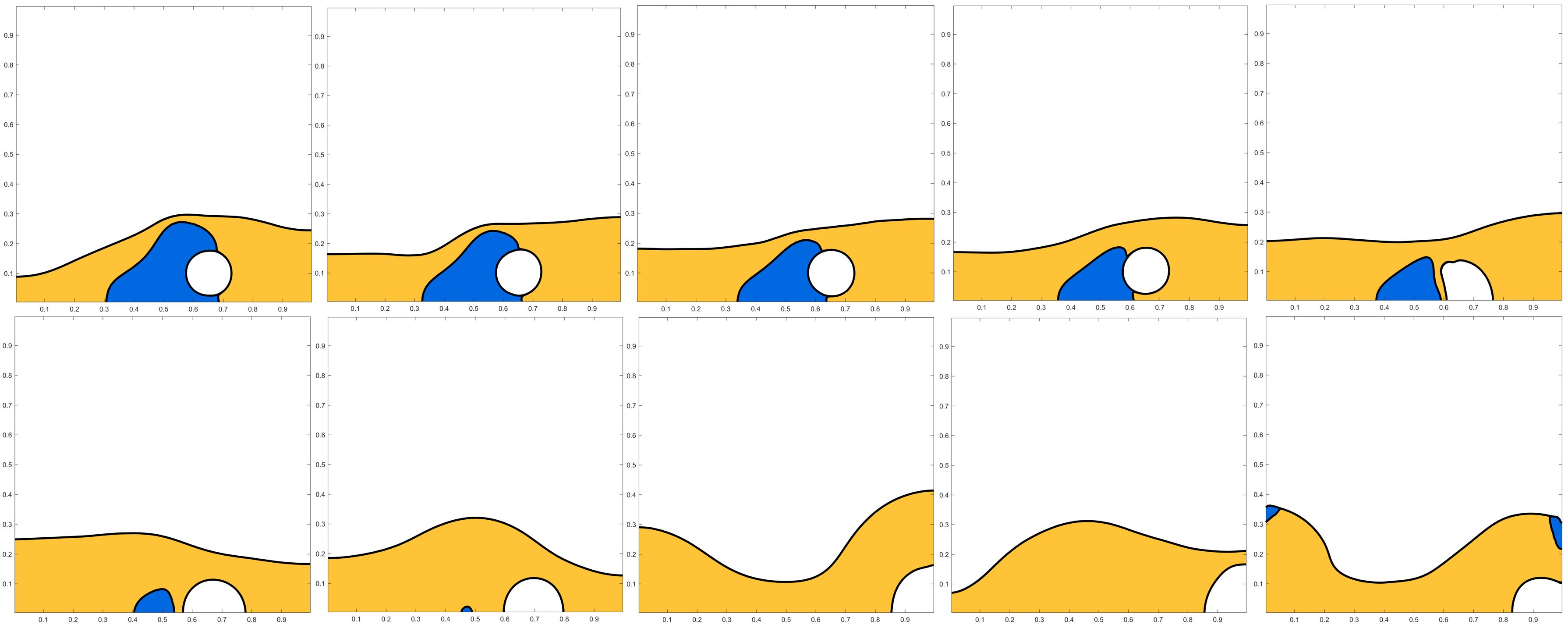}
	\includegraphics[scale=.43]{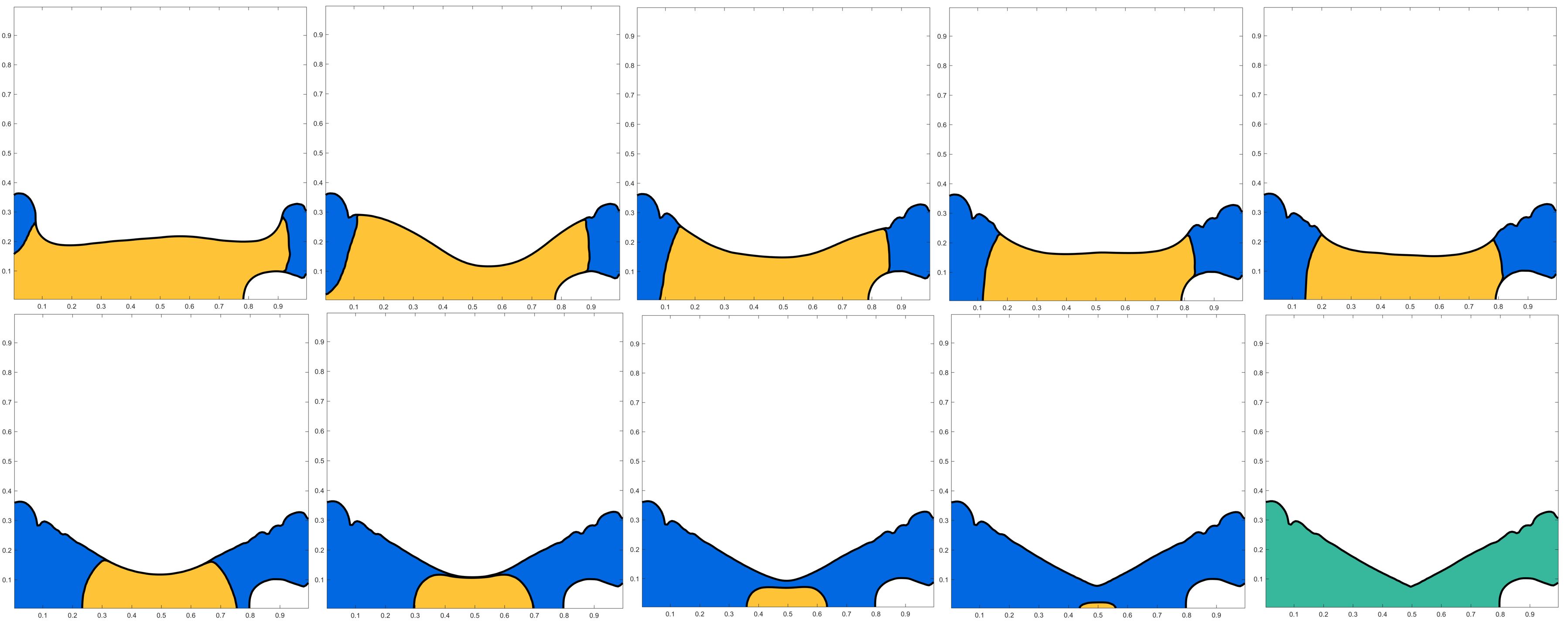}
	\caption{Results of melting and solidification. White: the gas phase, Orange: the liquid phase, Blue: the solid phase, Green: the solid phase when the phase change is finished. From left to right and top to bottom, $t=0.00$, $t=0.05$, $t=0.10$, $t=0.15$, $t=0.20$, $t=0.30$, $t=0.40$, $t=0.50$, $t=0.60$, $t=0.70$, $t=0.80$, $t=0.90$, $t=1.00$, $t=1.10$, $t=1.20$, $t=1.40$, $t=1.60$, $t=1.80$, $t=2.00$, $t=2.10$, $t=2.20$, $t=2.30$, $t=2.40$, $t=2.50$, $t=2.60$, $t=3.00$, $t=3.25$, $t=3.50$, $t=3.75$, and $t=4.00$. \label{Fig Square}}
\end{figure}

Here, we consider melting a solid rectangle and solidifying it again. The material properties and setup are identical to those in Section \ref{Sec Bubbles}, except that the thermal conductivity of the gas is $\kappa_G=100 \mathrm{W/(m \cdot K)}$ and that the bottom wall becomes adiabatic, and the temperature at the other boundaries is $2$ before $t=2$ then $0.5$. The initial condition of the phases is illustrated in the first snapshot of Fig.\ref{Fig Square}. A rectangular solid with a width $0.6$ and height $0.4$ is sitting above the bottom wall, and it traps two circular gas bubbles whose radii are $0.075$ and $0.1$, and centers are located at $(0.65,0.1)$ and $(0.35,0.25)$, respectively. The initial temperature is $0.8$ inside the solid rectangle including the gas bubbles, while it is $2$ elsewhere. Recall that the non-dimensionalized melting temperature is $1$.

Results are shown in Fig.\ref{Fig Square}. The two top corners of the solid rectangle first melt, and then the lateral edges. The gas in the larger bubble is released to the ambient, and the melted liquid covers the solid and flows downward to the bottom wall. The solid phase gradually disappears and the smaller gas bubble is finally released. The smaller bubble slides on the bottom wall back and forth, following the capillary wave above, and finally reaches the right wall. As the temperature at the boundaries becomes lower than the melting temperature, solidification first appears at the lateral walls, and the front of the liquid-solid interface moves towards the middle, along with the capillary wave moving up and down. At the end of the simulation, the melted liquid completely solidifies with the smaller gas bubble trapped at the bottom-right corner.

\section{Conclusion and future work}\label{Sec Conclusions}
In the present work, we consider the thermo-gas-liquid-solid flows, where the liquid and solid phases are experiencing solidification/melting. A novel consistent and conservative Phase-Field model is developed for such a kind of problem. 
The ingredients of the proposed model are the consistent and conservative Phase-Field method for incompressible two-phase flows \citep{Huangetal2020}, the fictitious domain Brinkman penalization (FD/BP) method for fluid-structure interactions \citep{Angotetal1999,BergmannIollo2011}, and the Phase-Field model of solidification in \citep{Boettingeretal2002}. These successful models are physically coupled using the \textit{consistency of reduction}, \textit{consistency of volume fraction conservation}, \textit{consistency of mass conservation}, and \textit{consistency of mass and momentum transport}. These consistency conditions, which have been successfully applied in isothermal, multiphase, multicomponent, immiscible, and incompressible flows \citep{Huangetal2020,Huangetal2020CAC,Huangetal2020N,Huangetal2020B,Huangetal2020NPMC}, are used in problems having variable temperature and phase changes for the first time, and are demonstrated to play an essential role in the present study. 
The Cahn-Hilliard equation Eq.(\ref{Eq Cahn-Hilliard}) is applied to locate the gas and phase change material.
The phase change equation Eq.(\ref{Eq Phase change}) is derived from the solidification model in \citep{Boettingeretal2002} using the diffuse domain approach \citep{Lietal2009}. Then the \textit{consistency of volume fraction conservation} is applied to admit the fully liquid/solid-state of the phase change material. The interpolation function in \citep{Boettingeretal2002} is also modified so that the equilibrium states of the order parameter in the model depend on the temperature in a physical sense, which resolves the issue of initiating the phase change when the phase change material is fully liquid/solid at the beginning. 
After applying the \textit{consistency of mass conservation}, we not only obtain the consistent mass flux, which appears in the momentum equation following the \textit{consistency of mass and momentum transport}, but also the divergence of the velocity Eq.(\ref{Eq Divergence}), which quantifies the volume change induced by the solidification/melting. 
Isothermal (or temperature equilibrium) solutions are admissible by the proposed energy equation Eq.(\ref{Eq Energy}) when the phase change is absent, after incorporating the \textit{consistency of mass conservation} and the \textit{consistency of volume fraction conservation}.
To confine the surface tension effect on the gas-liquid interface only, we propose a new continuous surface tension force based on the one in \citep{Huangetal2020}. The Carman-Kozeny equation \citep{Carman1997} is modified to enforce zero velocity in the solid phase. These two additional forces are added to the momentum equation Eq.(\ref{Eq Momentum}).
The proposed model defines the volume fractions of the gas, liquid, and solid phases unambiguously, and the volume change due to solidification/melting is included. The mass and energy conservation is always true, while the momentum conservation is honored if the solid phase is absent due to the no-slip condition at the solid boundary. The proposed model also satisfies the Galilean invariance. Moreover, we show in Theorem~\ref{Theorem Two-Phase}, Theorem \ref{Theorem FSI}, and Theorem \ref{Theorem Solidification} that the proposed model will automatically recover the corresponding two-phase models in \citep{Huangetal2020,Angotetal1999,BergmannIollo2011,Boettingeretal2002} when one of the phases is locally absent. 

The proposed model is numerically solved with a scheme that reproduces the physical connections in the model on the discrete level, and various numerical tests have been performed to verify and demonstrate the proposed model. 
Theorem \ref{Theorem Two-Phase}, Theorem \ref{Theorem FSI}, and Theorem \ref{Theorem Solidification} are verified with the large-density-ratio advection, the Couette flow, and the Stefan problem, whose exact solutions are available. The results from the proposed model not only agree very well with the exact solutions but also match the expectations from the theorems.
The volume change resulting from the phase change is illustrated, and it is quantitatively demonstrated to be consistent with the mass conservation. This physical behavior has not been captured in many existing models by assuming the divergence-free velocity all the time. The numerical error of mass conservation is very small, and therefore the present scheme conserves the mass satisfactorily, although not exactly. 

We illustrate that the surface tension force can affect the results significantly, especially when the velocity in the solid phase is not reduced to zero effectively, and the proposed surface tension force successfully confines its effective region at the gas-liquid interface and produces stable solutions. We also demonstrate that defining a solid viscosity much larger than the liquid one is not adequate to stop the movement of the solid phase. A new criterion, based on the scaling of the discretized inertial and viscous forces, is proposed. The solid viscosity needed to stop the solid motion is usually too large to obtain a stable solution. Therefore, increasing the solid viscosity is not an effective way to enforce zero velocity in the solid phase, and adding a drag force is preferable.
In addition to the proposed drag force, an alternative modification on the Carman-Kozeny equation \citep{Carman1997} is studied. Both the proposed one and the alternative one effectively leave to zero velocity in the solid phase and produce similar results, while the effective region of the alternative one is larger. We also analyze the scaling of the discretized inertial and viscous forces to determine the parameter $C_d$ in the drag force model. 

After verifying the proposed model, a comparison to experimental and other numerical data is conducted, and the proposed model produces results that agree well with those data. We discover that the Gibbs-Thomson and linear kinetic coefficients, i.e., $\Gamma_\phi$ and $\mu_\phi$, need to be carefully selected to obtain a quantitative agreement. $\mu_\phi$ and $\Gamma_\phi$ is positively correlated to the speed of phase change and the interface thickness, respectively, which can be explained by the energy mechanism of the Phase-Field model of solidification in \citep{Boettingeretal2002}. In practice, calibration may be needed to determine $\mu_\phi$, and we suggest that $\mu_\phi$ and $\Gamma_\phi$ take the same value. 
Finally, two challenging problems, including a wide range of material properties and strong interactions among different phases, are set up and successfully solved, which illustrates the capability of the model.

The present study proposes a practical framework to incorporate the solidification/melting of a pure material into liquid-gas flows. This method can be extended to include more complicated physics, e.g., the thermo-capillary effect, anisotropy or dendritic growth, and solute transport during the solidification. 
One may notice that the proposed surface tension force in Eq.(\ref{Eq Momentum}) for the thermo-gas-liquid-solid flows can be written as $\mathbf{f}_s=\phi \mathbf{f}_s'$, where $\mathbf{f}_s'$ denotes the original surface tension model for two-phase flows without considering the solid phase, and the present study uses the Phase-Field formulation, i.e., $\mathbf{f}_s'=\xi_\varphi \nabla \varphi$. Therefore, a possible way to include the thermo-capillary effect in the proposed model is to choose $\mathbf{f}_s'$ which incorporates that effect in two-phase flows, such as the one in \citep{Liuetal2014}. Another interesting but also practical direction to extend the proposed model is to include discrete particulate materials, due to their frequent appearance in industrial processes, like selective laser melting (SLM) which is an additive manufacturing (AM) method. Different from freezing the solid motion in the present study, the discrete particulate materials are allowed to move, driven by interaction forces between the particles and the fluids. A possible strategy is to follow the recent development in \citep{YuZhao2021}, where the interaction forces are provided in detail and the discrete particulate materials are modeled by the Discrete Element Method (DEM) to update their locations and velocities. The developed scheme preserves many physical properties of the proposed model on the discrete level, which helps to reduce the interference from numerical errors and therefore is preferred in the present study for verification and demonstration purposes. However, efficiency is less considered, and parallelization has not been implemented. Because of that, only two-dimensional results are presented, although both the model and scheme can be directly extended to three-dimensional problems. Efficiency becomes a critical issue when implementing the present method to practical problems, since those problems are usually three-dimensional and require long-time simulations. Although the developed scheme is decoupled and solves only linear systems, which is favorable for efficiency, the coupling among the governing equations of the proposed model is still strong. Developing an efficient parallelization strategy that honors the physical connections in the proposed model is an ongoing research. Adaptive grid refinement and time stepping are attractive directions to improve numerical simulations, while limiting computational cost. However, the physical properties of the model need to be preserved on the discrete level to avoid unphysical behaviors. This is a non-trivial problem and deserves further investigation.

\section*{Acknowledgments}
A.M. Ardekani would like to acknowledge the financial support from the National Science Foundation (CBET-1705371). This work used the Extreme Science and Engineering Discovery Environment (XSEDE) \cite{Townsetal2014}, which is supported by the National Science Foundation grant number ACI-1548562 through allocation TG-CTS180066  and TG-CTS190041.
G. Lin would like to acknowledge the support from National Science Foundation (DMS-1555072 and DMS-1736364, CMMI-1634832 and CMMI-1560834), and Brookhaven National Laboratory Subcontract 382247, ARO/MURI grant W911NF-15-1-0562, and U.S. Department of Energy (DOE) Office of Science Advanced Scientific Computing Research program DE-SC0021142.

\bibliographystyle{plain}
\bibliography{refs.bib}

\end{document}